\documentclass[journal]{aa}

\usepackage[dvips]{graphicx}
\usepackage{amsmath}
\usepackage{amssymb}
\usepackage{verbatim}
\usepackage{array}
\usepackage{txfonts}
\usepackage{natbib}

\bibpunct{(}{)}{;}{a}{}{,}

\newcommand{\celltspace}{\rule{0pt}{2.8ex}}
\newcommand{\cellbspace}{\rule[-1.4ex]{0pt}{0pt}}

\defcitealias{Martin:2009a}{Paper I}
\defcitealias{Limongi:2006}{LC06}
\defcitealias{Palacios:2005}{MMP05}
\defcitealias{Meynet:1997}{MM97}
\defcitealias{Woosley:1995}{WW95}
\defcitealias{Woosley:1995a}{WLW95}
\defcitealias{Woosley:2007}{WH07}

\begin{document}

% Title, header, abstract
\title{Predicted gamma-ray line emission from the Cygnus complex}
\titlerunning{Predicted gamma-ray line emission from the Cygnus complex}
\author{Pierrick Martin \inst{1,2} \and J\"urgen Kn\"odlseder \inst{1,2} \and Georges Meynet \inst{3} \and Roland Diehl \inst{4}}
\institute{Centre d'Etude Spatiale des Rayonnements, CNRS/UPS, 9, avenue colonel Roche, BP44346, 31028 Toulouse cedex 4, France  \and Universit\'e de Toulouse (UPS), Centre National de la Recherche Scientifique (CNRS), UMR 5187, France \and Geneva Observatory, CH-1290 Sauverny, Switzerland  \and Max Planck Institut f\"ur extraterrestrische Physik, Postfach 1312, 85741 Garching, Germany}
\date{Received: 10 July 2009 / Accepted: 17 November 2009}
\abstract{The Cygnus region harbours a huge complex of massive stars at a distance of 1.0-2.0\,kpc from us. About 170 O stars are distributed over several OB associations, among which the Cyg OB2 cluster is by far the most important with about 100-120 O stars. These massive stars inject large quantities of radioactive nuclei into the interstellar medium, such as $^{26}$Al and $^{60}$Fe, and their gamma-ray line decay signals can provide insight into the physics of massive stars and core-collapse supernovae.}{Past studies of the nucleosynthesis activity of Cygnus have concluded that the level of $^{26}$Al decay emission as deduced from CGRO/COMPTEL observations was a factor 2-3 above the predictions based on the theoretical yields available at that time and on the observed stellar content of the Cygnus region. We reevaluate the situation from new measurements of the gamma-ray decay fluxes with INTEGRAL/SPI (presented in a previous paper) and new predictions based on recently improved stellar models.}{We built a grid of nucleosynthesis yields from recent models of massive stars. Compared to previous works, our data include some of the effects of stellar rotation for the higher mass stars and a coherent estimate of the contribution from SNIb/c. We then developed a population synthesis code to predict the nucleosynthesis activity and corresponding decay fluxes of a given stellar population of massive stars.}{The observed decay fluxes from the Cygnus complex are found to be consistent with the values predicted by population synthesis at solar metallicity; and yet, when extrapolated to the possible subsolar metallicity of the Cygnus complex, our predictions fail to account for the INTEGRAL/SPI measurements. The observed extent of the 1809\,keV emission from Cygnus is found to be consistent with the result of a numerical simulation of the diffusion of $^{26}$Al inside the superbubble blown by Cyg OB2.}{Our work indicates that the past dilemma regarding the gamma-ray line emission from Cygnus resulted from an overestimate of the 1809\,keV flux of the Cygnus complex, combined with an underestimate of the nucleosynthesis yields. Our results illustrate the importance of stellar rotation and SNIb/c in the nucleosynthesis of $^{26}$Al and $^{60}$Fe. The effects of binarity and metallicity may also be necessary to account for the observations satisfactorily.}

\maketitle

% Introduction
\section{Introduction}
\label{introduction}

Conventionally defined as stars with initial masses above 8-10 and up to about 120\,M$_{\odot}$, massive stars play a fundamental role in astrophysics because of the very powerful behaviour they exhibit during their lives, from the early main-sequence to their final explosion (and even beyond thanks to the compact objects they leave behind). Usually clustered in OB associations of a few tens to several thousand members, massive stars severely affect their environment on the hundred-parsec scale through their sustained mechanical and radiative energy output: strong luminosity in the Lyman continuum, very energetic winds and mass loss episodes, and supernova explosions. In addition to mechanical energy, these stellar outflows also carry the products of nuclear burning and thus enrich the interstellar medium (ISM) with heavy elements. The physical and chemical evolution of the Galaxy is driven to a considerable extent by massive stars and a wide variety of Galactic objects follow from their evolution: supernovae (SNe) and their remnants (SNRs), neutron stars or stellar-mass black holes, some of which manifest in the form of pulsars or magnetars, X-ray binaries, or microquasars.\\
\indent A thorough understanding of the processes that rule the evolution of massive stars and their death as core-collapse SNe is therefore inevitable in astrophysics. Amongst the various observational approaches employed to explore the physics of massive stars and their SNe, gamma-ray line astronomy can be a very valuable tool. By studying the characteristic decay radiation of certain radio-nuclides produced in the hydrostatic and/or explosive burning stages of massive stars, gamma-ray line astronomy provides indirect access to the physical conditions that governed their synthesis and ejection in the ISM. Among the few radio-isotopes accessible to the present-day gamma-ray telescopes, the long-lived $^{26}$Al and $^{60}$Fe hold potential for interesting constraints on the evolution of massive stars.\\
\indent The Galactic decay emission from $^{26}$Al was extensively studied in the 1990s with the CGRO/COMPTEL instrument and the resultant cartography of the $^{26}$Al decay signal at 1809\,keV revealed that most of the emission comes from the inner Galaxy and from a few nearby star-forming regions \citep{Diehl:1994,Del-Rio:1996,Pluschke:2001}. The intensity distribution observed with COMPTEL allowed it to be established that massive stars are the dominant source of $^{26}$Al in the Galaxy \citep{Prantzos:1996,Knodlseder:1999}. In the early 2000s, the first detection of the Galactic $^{60}$Fe decay signal at 1173 and 1332\,keV was achieved from RHESSI observations \citep{Smith:2004}, with a measured flux of about 15\% of the $^{26}$Al decay flux. Although the decay emission was too weak to achieve a detailed cartography and thus confirm that massive stars are the main source of $^{60}$Fe, the combined analysis of $^{26}$Al and $^{60}$Fe decay signals appeared as a potentially strong constraint on the processes that govern the nucleosynthesis in massive stars and SNe.\\
\indent The launch in 2002 of the INTEGRAL gamma-ray observatory, equipped with the SPI high-resolution spectrometer, has opened the possibility of extending the analysis of the gamma-ray line emission from the Galaxy and add spectrometric information to the COMPTEL results. Both the $^{26}$Al and $^{60}$Fe decay signals have now been unambiguously detected by SPI on the Galactic scale \citep{Diehl:2006,Wang:2007}. The stronger $^{26}$Al emission made it possible to analyse its morphology in more detail and appeared particularly prominent in some specific star-forming regions of the Galaxy, in agreement with the COMPTEL observations. In a previous work \citep[][hereafter Paper I]{Martin:2009a}, we showed that the Cygnus region exhibits a relatively strong signal of $^{26}$Al decay, so we investigated further the gamma-ray line emission of both $^{26}$Al and $^{60}$Fe from Cygnus.\\
\indent The Cygnus region and its peculiarities were presented in detail in \citetalias{Martin:2009a} and we only repeat the main characteristics of the region here and the principal arguments for studying it in nuclear gamma-ray lines. The Cygnus region, which we consider to be that part of the Galactic disk located between longitudes 70$^{\circ}$ and 90$^{\circ}$, harbours a very impressive concentration of massive stars at only $\sim$ 1-2\,kpc from us. About 170 O stars have been recorded \citep{Knodlseder:2002}, distributed in 6 OB associations and a dozen open clusters, among which the spectacular Cyg OB2 cluster that alone comprises 100-120 O stars \citep{Knodlseder:2000,Comeron:2002}. In the following, we collectively refer to these OB associations and open clusters (the details of which are presented later) as the ``Cygnus complex". With this term, we aim at gathering all these rich and nearby stellar groups that are relatively correlated in terms of distance (between 1 and 2\,kpc) and thus contribute to the specificity of the Cygnus region. This ``Cygnus complex" appears to be quite young, as shown by the lack of radio supernova remnants (SNRs) or pulsars. The Green catalogue\footnote{Green D.A., 2006, "A Catalogue of Galactic Supernova Remnants (2006 April version)", Astrophysics Group, Cavendish Laboratory, Cambridge, UK (http://www.mrao.cam.ac.uk/surveys/snrs/).} gives about 10 radio SNRs between 70$^{\circ}$ and 90$^{\circ}$, but according to a compilation of distances by \citet{Kaplan:2004}, none of these objects can be related to the Cygnus complex. \citet{Wendker:1991} concluded from the sensitivity of their radio survey of the region that no other radio SNR exists up to a distance of 10\,kpc. Additional evidence for the relative youth of the complex comes from the absence of radio pulsars towards Cygnus up to 4\,kpc, as indicated by the ATNF catalogue\footnote{http://www.atnf.csiro.au/research/pulsar/psrcat/} \citep{Manchester:2005}, although recent observations with the LAT telescope aboard the Fermi Gamma-Ray Space Telescope have led to the discovery of the young energetic pulsar PSR J2032+4127 \citep{Abdo:2009d} that is likely associated with Cyg OB2 \citep{Camilo:2009}. For those reasons, the Cygnus region appears well-suited to studying massive stars in their hydrostatic phases.\\
\indent The Cygnus population of massive stars is close and rich enough to ensure sufficient gamma-ray line fluxes and therefore constitutes a good opportunity to test our understanding of the structure and evolution of massive stars, as revealed by their nucleosynthetic activity. A number of studies have been devoted to the nucleosynthesis of $^{26}$Al and $^{60}$Fe in the Cygnus region \citep[see for example][]{Del-Rio:1996,Pluschke:2000,Cervino:2000,Pluschke:2002,Knodlseder:2002} and the latest studies agree that the level of the 1809\,keV emission from Cygnus as deduced from CGRO/COMPTEL observations is a factor of 2-3 above the predictions, when the latter are based on the theoretical yields by \citet{Meynet:1997}, \citet{Woosley:1995} and \citet{Woosley:1995a} and on the observed stellar content of the Cygnus region \citep[including the upward revision of the stellar population of Cyg OB2 by][]{Knodlseder:2000}. \citet{Pluschke:2000} show that the discrepancy could be alleviated if one includes the enhanced yields from massive close binary systems in the calculation. Alternatively, \citet{Pluschke:2002} argue that the problem may well come from an underestimate of the stellar content of the Cygnus OB associations due to a strong interstellar extinction, as was shown to be the case for Cyg OB2. In \citet{Knodlseder:2002}, the authors speculate that future stellar models, especially those that include the effects of stellar rotation, could solve the dilemma.\\
\indent In \citetalias{Martin:2009a}, we used the INTEGRAL/SPI observations to reassess the $^{26}$Al and $^{60}$Fe decay emission from the Cygnus region. In particular, we disentangled in a coherent way the contribution of the nearby clustered OB population of the Cygnus complex from the contribution of the more diffuse and spatially extended non-clustered population, which is likely to contribute as well to the observed emission. We now compare our INTEGRAL/SPI observations of the gamma-ray line emission of the Cygnus complex with theoretical predictions based on the latest models of stellar nucleosynthesis. In recent years, stellar models have been improved with modifications, such as the inclusion of stellar rotation or the revision of the mass loss rates and nuclear cross-sections. These upgrades have already proven to bring simulations closer to observations. For instance, models including rotation can account for changes in surface abundances occurring during the main sequence phases \citep{Hunter:2008,Maeder:2009} and for the observed variations of the number of Wolf-Rayet to O-type stars as a function of metallicity in constant star formation regions \citep{Meynet:2005}. In the following, we show that the new stellar models, combined with the work done on observations in \citetalias{Martin:2009a}, considerably alleviate the long standing discrepancy between theoretical and observed decay fluxes from Cygnus.

% Observational constraints from INTEGRAL/SPI
\section{Observational constraints from INTEGRAL/SPI}
\label{observations}

In \citetalias{Martin:2009a}, we presented the analysis of about 4 years of INTEGRAL/SPI observations amounting to a total effective exposure time of 63.2\,Ms. Most of this exposure is confined to the Galactic plane, with 10.8\,Ms specifically covering the Cygnus region. In this section, we review the main results that were obtained on the $^{26}$Al and $^{60}$Fe decay emission.

% The 1809keV emission from 26Al
\subsection{The 1809\,keV emission from $^{26}$Al}
\label{obs_26Al}

\indent The $^{26}$Al isotope is produced predominantly by massive stars at various stages of their existence and it is released in the ISM by both the stellar winds (of Wolf-Rayet stars especially) and the SNe. Nucleosynthesis of $^{26}$Al follows from the $^{25}$Mg(p,$\gamma$)$^{26}$Al reaction, which mostly occurs during H central-burning (with $^{25}$Mg coming from initial metallicity), C/Ne shell-burning (with $^{25}$Mg coming from the CNO initial metallicity through the $^{14}$N($\alpha$,$\gamma$)$^{18}$F(e$^-$)$^{18}$O($\alpha$,$\gamma$)$^{22}$Ne($\alpha$,n)$^{25}$Mg chain) and explosive burning of the C-shell (with $^{25}$Mg coming from $^{24}$Mg(n,$\gamma$)$^{25}$Mg, $^{24}$Mg being a product of C/Ne burning).\\
\indent The mean $^{26}$Al lifetime of about 1\,Myr allows it to diffuse in the local ISM before decaying into $^{26}$Mg. As a consequence, the corresponding 1809\,keV emission is of diffuse nature with an overall distribution that follows that of massive stars in our Galaxy, as also traced for instance by the microwave free-free emission from HII regions or the infrared emission from heated dust \citep{Knodlseder:1999}. Most of the Galactic 1809\,keV flux is thus confined to the Galactic plane and inside the inner regions, except for a notable emission from the Cygnus and Sco-Cen regions (see \citetalias{Martin:2009a}).\\
\indent From the INTEGRAL/SPI data, we have found that the diffuse 1809\,keV emission from the Cygnus region is well represented by a $3^{\circ} \times 3^{\circ}$ 2D Gaussian intensity distribution (which implies a characteristic angular size of 9-10$^{\circ}$) centred approximately on the position of the Cyg OB2 cluster. This confirms expectations that the latter dominates the energetics and nucleosynthetic activity in this area. It should be emphasised, however, that we only have weak constraints on the maximal extent of the emission due to intrinsic limitations of the coded mask imaging system of the SPI instrument and yet, the size we found is consistent with the results obtained from the CGRO/COMPTEL observations \citep{Pluschke:2001}. The total flux from the extended source in the Cygnus direction is (6.0 $\pm$ 1.0) $\times$ 10$^{-5}$\,ph\,cm$^{-2}$\,s$^{-1}$, where (3.9 $\pm$ 1.1) $\times$ 10$^{-5}$\,ph\,cm$^{-2}$\,s$^{-1}$ is attributed to the Cygnus complex and the remaining flux stems from Galactic disk foreground and background emissions.\\
\indent Thanks to the high spectral resolution of the SPI instrument, we were able to study the profile of the 1809\,keV line. We observed a line position corresponding to a radial velocity of 33 $\pm$ 66\,km\,s$^{-1}$, which agrees with the overall radial motions in the direction and at the distance of the Cygnus complex, as indicated for instance by spectrometric observations of the CO line \citep{Dame:2001}. We also measured a slight line broadening that, within the statistical uncertainty, is consistent with $^{26}$Al being still or in a medium with turbulent or expansion velocities of up to 100-200\,km\,s$^{-1}$.

% The 1173/1332keV emission from 60Fe
\subsection{The 1173 and 1332\,keV emission from $^{60}$Fe}
\label{obs_60Fe}

\indent The massive star origin of the $^{60}$Fe isotope is far less established than for $^{26}$Al. The nucleosynthesis of $^{60}$Fe by massive stars was addressed by \citet{Timmes:1995} and extensively reviewed by \citep{Limongi:2006}. From these works, we know that massive stars are a source of $^{60}$Fe through the $^{59}$Fe(n,$\gamma$)$^{60}$Fe reaction that occurs during He shell-burning, C shell-burning, and explosive burning of the C-shell. In these processes, the seed amounts of Fe come from the initial metallicity and so do the neutrons, through the $^{22}$Ne($\alpha$,n)$^{25}$Mg reaction (with $^{22}$Ne coming from the initial CNO, see \ref{obs_26Al}). The $^{60}$Fe thus produced by massive stars is then released in the ISM only through SN explosions and no wind contribution is expected. Other objects such as AGB stars \citep{Karakas:2007} or SNe of type Ia \citep{Iwamoto:1999} are also potential producers of $^{60}$Fe. Although the relative contribution of each class of objects remains entirely unconstrained on the observational level, the available models for the various potential sources indicate that core-collapse SNe should provide most of the Galactic $^{60}$Fe.\\
\indent The observed intensity of the Galactic decay emission at 1173 and 1332\,keV is a factor of 7-8 lower than the 1809\,keV emission from $^{26}$Al \citep[][and \citetalias{Martin:2009a}]{Harris:2005,Wang:2007}, so the detection of any $^{60}$Fe signal from Cygnus may at first sight be difficult given the sensitivity of INTEGRAL/SPI. In addition, the Cygnus complex seems quite young, and it may well be that its stellar population has so far not produced a single SN, which would mean no release of $^{60}$Fe into the ISM, hence no 1173/1332\,keV signal from the Cygnus complex at all. As a matter of fact, we did not observe any 1173/1332\,keV emission from the Cygnus direction and derived a 2$\sigma$ upper limit on the $^{60}$Fe decay flux of 1.6 $\times$ 10$^{-5}$\,ph\,cm$^{-2}$\,s$^{-1}$ \citep[consistent with the value obtained by][]{Wang:2007}.

% Population synthesis
\section{Population synthesis}
\label{popsim}

\indent In this section we present the most important aspects of the population synthesis simulations we performed to compare our INTEGRAL/SPI observations with. In this effort, we took advantage of the recent progress by two research groups in the field of massive stars nucleosynthesis. In the following, we introduce the specificities of each grid of stellar models and provide some detail about their implementation in our predictive tool.

% The stellar models
\subsection{The stellar models}
\label{popsim_models}

\indent Once the massive stars had been recognised as the dominant source of the Galactic $^{26}$Al \citep{Prantzos:1996,Knodlseder:1999}, the question of its origin shifted to identifying the subset of massive stars and/or evolutionary phases responsible for its ejection. For historical reasons, mostly linked to the evolution of the stellar models, two ``competing" sources were discussed: stars less massive than about 35\,M$_{\odot}$ through their SNII explosions, and stars more massive than 35\,M$_{\odot}$ through their WR winds \citep[see the review by][hereafter LC06]{Limongi:2006}. The debate then settled on the relative contribution of SNII and WR to the Galactic $^{26}$Al budget and, in this situation, the $^{60}$Fe appeared as a promising discriminating agent since it is released only by SN explosions.\\
\indent The work of \citetalias{Limongi:2006} has broadened the issue thanks to a new grid of stellar models. The authors computed the nucleosynthesis yields for a wide range of non-rotating solar metallicity models, with initial stellar masses from 11 to 120\,M$_{\odot}$. The models were followed over their entire life, including both the hydrostatic phases and the explosive burning episodes of the supernova explosion. The main outcomes of the \citetalias{Limongi:2006} study are:
\begin{enumerate}
\item Over the mass interval considered, most of the ejected $^{26}$Al has explosive origins. For the particular case of the most massive stars, typically the WR progenitors, the production of $^{26}$Al due to explosive burning of the C/Ne shells exceeds the quantity synthesised by central H-burning and carried away by the stellar winds.
\item The production of $^{60}$Fe (by C/Ne convective shells burning mostly) of the most massive stars that explode as SNIb/c is up to a factor 10 higher than the yields associated with the lower mass stars that explode as SNII.
\end{enumerate}
It has always been clear that a dichotomy between WR and SNII as the sources of $^{26}$Al is inadequate and that the whole contribution of each massive star should be accounted for to understand the origin of both the $^{26}$Al and $^{60}$Fe. Moreover, the initial metallicity may have a strong impact on the relative contributions of WR winds and SN explosions to the production of $^{26}$Al, as will be illustrated later.\\
\indent In parallel to the work of \citetalias{Limongi:2006}, grids of stellar models including stellar rotation have revealed that the latter strongly affects the evolution and nucleosynthesis of massive stars. \citet{Meynet:2003,Meynet:2005} have computed the evolution of rotating models of stars with initial masses from 20 to 120\,M$_{\odot}$ up to the end of central He-burning. These rotating models were computed for velocities compatible with the mean observed velocities for OB stars, and the impact of stellar rotation on the production of $^{26}$Al has been examined by \citet{Palacios:2005}. The main outcomes of these works (hereafter collectively referred to as \citetalias{Palacios:2005}) are:
\begin{enumerate}
\item Rotation increases the convection and the circulation of chemical species inside the star. This implies a larger initial reservoir of $^{25}$Mg available for proton capture (in the case of non-zero initial metallicity) and an earlier ejection of the $^{26}$Al produced by central H-burning.
\item Rotation increases the mass loss by altering the surface gravity, the effective temperature, and the opacity of the envelope. This, combined with the more efficient diffusion of the burning products to the stellar surface, results in an earlier transition to the WR phase and a decrease in the minimum initial mass to become WR. As a result, the  $^{26}$Al contained in the He core (following central H burning) is released sooner and by more stars.
\end{enumerate}
In summary, stellar rotation alters the $^{26}$Al yields in two ways: the quantities ejected in the WR winds are increased and the time profile of the injection into the ISM is different as well. While the details of the $^{26}$Al release with time are not important for a large, stationary system such as the Galaxy, for which only the total yields per star are relevant, the Cygnus complex of massive stars is a young, evolving region and the comparison of observations with predictions therefore includes a temporal dimension. Therefore, a reliable population synthesis model should not only be able to reproduce the observed 1809\,keV flux from Cygnus but also reach an agreement within a time range that is consistent with the age estimates of the specific stellar clusters.\\
\indent Besides the above improvements, both grids of stellar models are also based on revised mass loss rates taking the effect of clumping in the winds into account and on revised nuclear reaction rates. \citetalias{Palacios:2005} also explored the influence of the initial metallicity $Z$ by computing four grids at $Z=0.004, 0.008, 0.02$, and 0.04. The grid of \citetalias{Limongi:2006} was only computed for a Z=0.02 solar metallicity.
\begin{figure}[t]
\begin{center}
\includegraphics[width=\columnwidth]{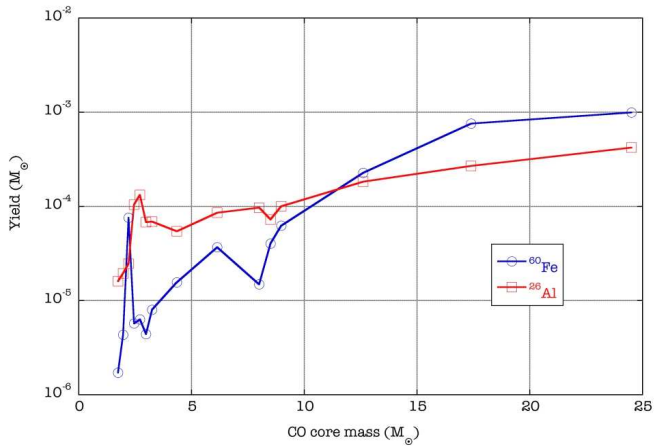}
\caption{Supernova yields of $^{26}$Al and $^{60}$Fe as a function of the CO core mass \citepalias[from the models of][]{Limongi:2006}. These curves are used to associate the purely hydrostatic models of \citetalias{Palacios:2005} with supernova yields.}
\label{fig_yieldsvsCOcore}
\end{center}
\end{figure}

% The grid of yields
\subsection{The grid of yields}
\label{popsim_grid}

\begin{figure}[t]
\begin{center}
\includegraphics[width=\columnwidth]{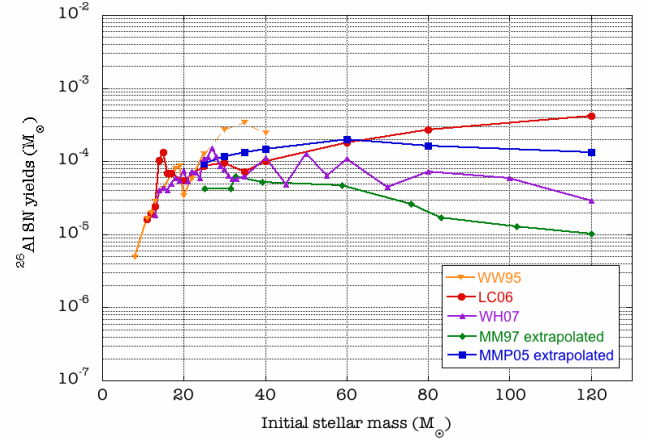}
\caption{$^{26}$Al supernova yields from \citetalias{Limongi:2006} and \citetalias{Woosley:2007}, compared to the older results from \citetalias{Woosley:1995}. Also plotted are the yields for \citetalias{Meynet:1997} and \citetalias{Palacios:2005} extrapolated through CO core mass from \citetalias{Woosley:1995a} and \citetalias{Limongi:2006} respectively.}
\label{fig_26AlSNyields}
\end{center}
\end{figure}

\begin{figure}[t]
\begin{center}
\includegraphics[width=\columnwidth]{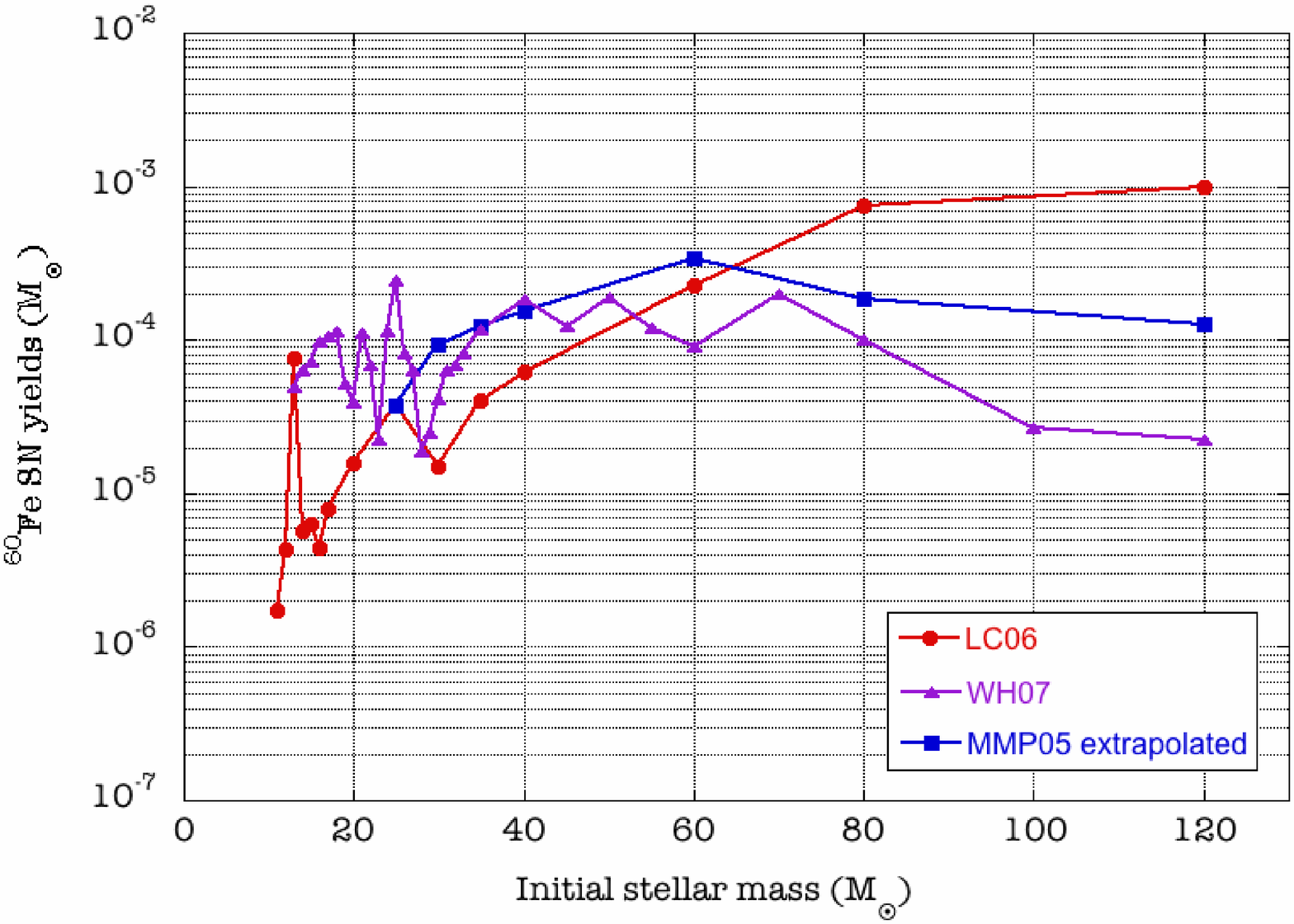}
\caption{$^{60}$Fe supernova yields from \citetalias{Limongi:2006} and \citetalias{Woosley:2007}. Also plotted are the yields for \citetalias{Palacios:2005} extrapolated through CO core mass from \citetalias{Limongi:2006}.}
\label{fig_60FeSNyields}
\end{center}
\end{figure}

The \citetalias{Limongi:2006} grid is complete and homogeneous but does not include stellar rotation and is only available for solar metallicity, while the \citetalias{Palacios:2005} grid includes the effects of stellar rotation and metallicity, but only covers the longest hydrostatic phases of those stars initially more massive than 20\,M$_{\odot}$. To make the most realistic predictions of the nucleosynthesis activity of the Cygnus complex, we therefore built our own grid of yields from the works of \citetalias{Limongi:2006} and \citetalias{Palacios:2005}.\\
\indent Before entering the description of our grid, we emphasise the terminology that will be used thereafter. We divided the nucleosynthesis contributions into ``wind yields" and ``SN yields", which is not equivalent to a separation into hydrostatic and explosive yields. Wind yields pertain to $^{26}$Al alone and are obviously of hydrostatic origin, whereas SN yields concern both isotopes and have both hydrostatic and explosive origins. For instance, the $^{60}$Fe released in a SN explosion comes from the hydrostatic burning of the He and C convective shells but also from the explosive burning of the C shell. We also note that the wind yield is a progressive, time-dependent outflow of $^{26}$Al, whereas the SN yield is considered an instantaneous release of $^{26}$Al and $^{60}$Fe.\\
\indent The grid we used in our population synthesis code is based on the following assumptions:
\begin{enumerate}
\item The lower mass stars, from 11 up to 20\,M$_{\odot}$, are represented by the models of \citetalias{Limongi:2006} for both their wind and SN contributions\footnote{In practice, however, the $^{26}$Al wind contribution of these stars is negligible because they experience little mass loss compared to the higher mass stars.}. It has also been assumed that these \citetalias{Limongi:2006} models are valid at metallicities other than solar. We will discuss this assumption later on.
\item The higher mass stars, from 20 to 120\,M$_{\odot}$ (typically those that become WR stars), are represented for their wind contribution by the rotating models of \citetalias{Palacios:2005} for the four metallicities considered by these authors. Each of these stellar tracks is then associated to one of the SN yields of \citetalias{Limongi:2006} through its CO core mass (as explained below).
\end{enumerate}
In the composition of our grid, we assumed that the SN yield is strongly linked to the final CO core mass of the model \citep[a hypothesis also used by][]{Cervino:2000}. This can first be justified by the fact that, at the moment of core-collapse, most of the $^{26}$Al and $^{60}$Fe synthesised so far is locked in the CO core as a result of the He/C/Ne shell-burning episodes. Second, it seems reasonable to consider that the amounts of $^{26}$Al and $^{60}$Fe produced explosively (through explosive burning of the C-shell) depend on the size of the CO core (even if other parameters, such as the chemical profile and the mass-radius relation, may be important). From the \citetalias{Limongi:2006} grid of models, it is possible to draw relations between the SN yields and the CO core masses (see Fig. \ref{fig_yieldsvsCOcore}). These relations are then used to associate the purely hydrostatic models of \citetalias{Palacios:2005} with SN yields through their final CO core mass that, for same initial stellar mass, differs from that obtained by \citetalias{Limongi:2006}. Indeed, for stars below about 60\,M$_{\odot}$, stellar rotation increases the final CO core mass because of an enhanced convection; above 60\,M$_{\odot}$, rotation causes a decrease in the CO core mass because of an enhanced mass loss \citep{Meynet:2003,Meynet:2005}.\\
\indent This approach takes into account how stellar rotation can affect the SN yields through its impact on the stellar evolution. It should still be emphasised that the SN yields used in our grid come only from non-rotating models. The impact of stellar rotation on the nucleosynthesis of the latest evolutionary stages remains unexplored up to now. \citetalias{Limongi:2006} pointed out how important the shell convection efficiency is to the synthesis of $^{26}$Al and $^{60}$Fe, but how stellar rotation affects the nucleosynthesis of the He/C/Ne shell-burning episodes is still an open question. Our limited knowledge on that point therefore constitutes a potential source of error for our SN yields.\\
\begin{figure}[t]
\begin{center}
\includegraphics[width=\columnwidth]{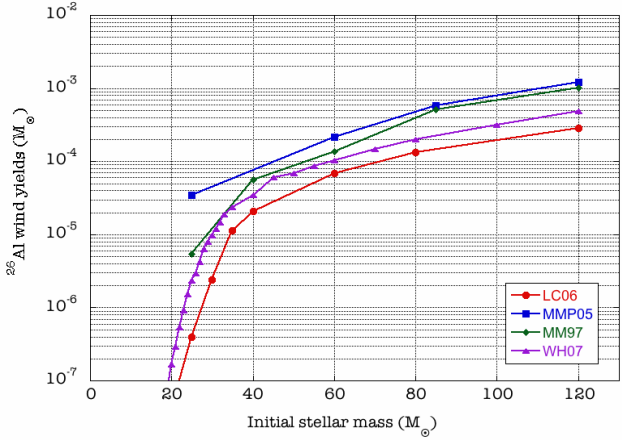}
\caption{$^{26}$Al wind yields from \citetalias{Limongi:2006}, \citetalias{Palacios:2005}, and \citetalias{Woosley:2007}, compared to the older results from \citetalias{Meynet:1997}.}
\label{fig_26Alwindyields}
\end{center}
\end{figure}
\indent In Figs. \ref{fig_26Alwindyields} and \ref{fig_26AlSNyields}, we show the $^{26}$Al yields for \citetalias{Limongi:2006} and \citetalias{Palacios:2005} (where the SN yields for \citetalias{Palacios:2005} were extrapolated through CO core mass, as explained above), compared to the previous generation of yields from \citet[][hereafter MM97]{Meynet:1997}, \citet[][hereafter WW95]{Woosley:1995} and \citet[][hereafter WLW95]{Woosley:1995a}. This previous generation of yields was used in a population synthesis analysis by \citet{Cervino:2000} and \citet{Knodlseder:2002}, who concluded that they could not account for the 1809\,keV flux from Cygnus observed by CGRO/COMPTEL \citep[see also][]{Pluschke:2001}. In Fig. \ref{fig_60FeSNyields}, we show the $^{60}$Fe yields for \citetalias{Limongi:2006} and \citetalias{Palacios:2005} (where again the SN yields for \citetalias{Palacios:2005} were extrapolated through CO core mass).\\
\indent In all three figures, we also compare the set of yields obtained by \citet[][hereafter WH07]{Woosley:2007}. These yields come from a large grid of stellar models (from 13 to 120\,M$_{\odot}$) that basically includes revised mass loss rates and revised nuclear reaction rates, but no rotation. The authors used an initial metallicity of Z=0.016, corresponding to the revised solar abundances of \citet{Lodders:2003}, and computed both hydrostatic and explosive phases. In its general features, the \citetalias{Woosley:2007} grid is very similar to the \citetalias{Limongi:2006} grid, except for the different initial metallicity. In the details, however, many aspects of stellar physics were implemented differently (such as the mass loss prescriptions, see below). As seen in the next paragraphs, the differences between the \citetalias{Woosley:2007} yields and the \citetalias{Limongi:2006} and \citetalias{Palacios:2005} yields illustrate the extent to which the implementation of the stellar physics can affect the nucleosynthesis results.\\
\indent In Fig. \ref{fig_26Alwindyields}, we see that the \citetalias{Palacios:2005} yields obtained from rotating models are more enhanced than the yields of \citetalias{Meynet:1997}, and this despite that the mass loss rates used in the \citetalias{Palacios:2005} models are lower than in the \citetalias{Meynet:1997} models. This illustrates the importance of the effects of rotation that, as explained in \ref{popsim_models}, favours the production of $^{26}$Al and its subsequent ejection by the stellar winds. The \citetalias{Woosley:2007} wind yields are all above the yields of \citetalias{Limongi:2006} (up to a factor of 5 in some cases) but always remain below the \citetalias{Palacios:2005} yields. This fact very likely comes from the absence of rotation in the \citetalias{Woosley:2007} models, but the difference between \citetalias{Woosley:2007} and \citetalias{Limongi:2006} could seem surprising. Indeed, a lower initial metallicity like Z=0.016 means that all other parameters being unchanged, a reduced initial reservoir of $^{25}$Mg available for proton capture, hence lower $^{26}$Al wind yields, but a thorough comparison of the \citetalias{Woosley:2007} and \citetalias{Limongi:2006} models is beyond the scope of this paper. The same rates were used for the main reactions governing $^{26}$Al production, but the higher mass models of \citetalias{Woosley:2007} were found to experience more mass loss (as discussed in more detail below). This may explain the situation above 40\,M$_{\odot}$. For the lower-mass models, however, the higher $^{26}$Al wind yields obtained with a lower initial metallicity may stem from the different treatment of the convective physics and/or to the different mass loss rate prescription during main sequence or red supergiant stages.\\
\indent In Fig. \ref{fig_26AlSNyields}, the $^{26}$Al SN yields of \citetalias{Limongi:2006} are plotted together with the SN yields extrapolated through the CO core mass for the \citetalias{Palacios:2005} rotating WR models. One clearly sees the effect of rotation on the size of the CO core, hence on the extrapolated SN yields, as discussed above. An older grid of yields is also plotted, which consisted of the \citetalias{Woosley:1995} SNII yields up to 25\,M$_{\odot}$ and the SNIb/c yields associated with the \citetalias{Meynet:1997} non-rotating WR models. For these SNIb/c yields, the yield versus CO core mass relation was derived from the pure He star models of \citetalias{Woosley:1995a}. Regarding the $^{26}$Al SN yields of \citetalias{Woosley:2007}, they are very similar to those of \citetalias{Limongi:2006} up to 40\,M$_{\odot}$ (variations remain within $\pm$ 50\%) except for the stars around 14-15\,M$_{\odot}$, for which \citetalias{Limongi:2006} found a dramatic increase of the $^{26}$Al SN yield not seen in \citetalias{Woosley:2007}. Beyond 40\,M$_{\odot}$, the yields of \citetalias{Woosley:2007} are below those of \citetalias{Limongi:2006} by a factor of up to 10 for the 120\,M$_{\odot}$ model. This is very likely an effect of the stronger mass loss experienced by the \citetalias{Woosley:2007} higher mass models.\\
\indent Again, no detailed comparison was performed but we noted that the \citet{Wellstein:1999} mass loss prescription (scaled by 1/3 to account for clumping) used by \citetalias{Woosley:2007} gives mass loss rates in the late WR stages (typically WNE and WCO stars) that are a factor of 2-3 above the \citet{Nugis:2000} prescription used by \citetalias{Limongi:2006}. This is obvious from the final presupernova masses of the 40-120\,M$_{\odot}$ models: while those obtained by \citetalias{Limongi:2006} are in the 12-20\,M$_{\odot}$ range, those obtained by \citetalias{Woosley:2007} are between 6-10\,M$_{\odot}$. It is interesting to note that \citetalias{Limongi:2006} tested the impact of increased mass loss during the WNE and WCO phases by using the prescription of \citet{Langer:1989}, which gives rates a factor of 2-3 above the rates of \citet{Nugis:2000}. In this case, their 40-120\,M$_{\odot}$ models ended up with final masses and $^{26}$Al SN yields that are very similar to those of \citetalias{Woosley:2007}. A stronger mass loss actually reduces the size of the He and CO cores and leads to lower $^{26}$Al SN yields. The difference between the \citetalias{Woosley:2007} and \citetalias{Limongi:2006} $^{26}$Al SN yields is all the more important because in the \citetalias{Woosley:2007} yields there is a contribution from the $\nu$-process that is absent in the \citetalias{Limongi:2006} yields (see \ref{predvsobs_fluxes_error}). The alternative grid of \citetalias{Woosley:2007} thus illustrates how important mass loss can be for nucleosynthesis.\\
\indent In Fig. \ref{fig_60FeSNyields}, the $^{60}$Fe yields of \citetalias{Limongi:2006} and \citetalias{Woosley:2007} are plotted with the SN yields extrapolated through the CO core mass for the \citetalias{Palacios:2005} rotating WR models. As for $^{26}$Al SN yields, the effect of rotation on the size of the CO core, hence on the extrapolated yields, is manifest. The \citetalias{Woosley:2007} yields are apparently quite different from those of \citetalias{Limongi:2006}. Between 10-20\,M$_{\odot}$ they are up to a factor of 20 above those of \citetalias{Limongi:2006}, then between 20-50\,M$_{\odot}$ they are still above those of \citetalias{Limongi:2006} by more moderate factors of up to 5, and beyond 50\,M$_{\odot}$ they drop below \citetalias{Limongi:2006} by quite large factors. For the higher mass models, the lower yields of \citetalias{Woosley:2007} very likely come from the stronger mass loss of WR stars, as already discussed above. Here again, the same effect was noted by \citetalias{Limongi:2006} when they used the increased mass loss rates of \citet{Langer:1989}. For the lower masses, however, the differences are not so easy to explain. The lower initial metallicity should have resulted, all other parameters being unchanged, in lower yields since $^{60}$Fe is a pure secondary product. The major reaction rates for the production and destruction of $^{60}$Fe were found to be identical in \citetalias{Woosley:2007} and \citetalias{Limongi:2006}, so the difference may come from another aspect of stellar physics such as convection or mixing.

% The stellar content of the Cygnus complex
\subsection{The stellar content of the Cygnus complex}
\label{popsim_stars}

\indent The last ingredient of our simulation is the definition of the stellar population to be modelled. The peculiarities of the Cygnus region have fostered a great number of studies that together provide us with a pretty good knowledge of the area. In particular, recent works have lead to a substantial revision of the stellar content of the Cyg OB2 cluster and thus allow for more accurate predictions.\\
\indent From the stellar census of \citet{Knodlseder:2000}, \citet{Knodlseder:2002}, and \citet{Le-Duigou:2002}, we selected the most relevant OB associations and open clusters of the Cygnus region. We restricted ourselves to a longitude range of 70-90$^{\circ}$ and among the stellar groups falling in that interval, we excluded the ones with ages above 20\,Myr (since their decay fluxes are negligible after that time, see Fig. \ref{fig_fluxvsIMF}). The selected objects are listed in Table \ref{tab_ OBasso}, together with their main characteristics. \citet{Knodlseder:2002} indicate that most open clusters are potentially physically related to some of the OB associations. In total, our selection amounts to a total of $\sim$ 170 O stars, most of which are in the Cyg OB2 cluster. This ensemble is what we have called so far the Cygnus complex. \citet{Schneider:2006} demonstrate that Cyg OB1, 2, and 9 very likely originated from the same giant molecular cloud, which gives a physical reality to this grouping. The connection of Cyg OB3, 7, and 8 to this complex is, however, still uncertain.\\
\indent Whether all these clusters share a common origin does not matter at all for our purpose. Our aim here is simply to gather the nearby clustered stellar population that is very likely responsible for most of the strong 1809\,keV emission from Cygnus. In some previous studies \citep{Del-Rio:1996,Pluschke:2002}, the total 1809\,keV flux from the Cygnus region was compared to theoretical predictions based on the clustered stellar content of the Cygnus region supplemented by a few isolated WR stars and SNRs. In our study, we focus both observationally and theoretically on the clustered stellar content alone, because the clustered population is likely to be better inventoried than the isolated or non-clustered population. In \citetalias{Martin:2009a}, we used the INTEGRAL/SPI data to disentangle the 1809\,keV flux of the Cygnus complex from the more spatially extended emission likely due to unclustered objects like non-member WR and OB stars (runaway stars or stars from previous star formation episodes, see \citet{Comeron:2007} and \citet{Comeron:2008}) or isolated SNRs. In the present paper, we compare this flux to population synthesis predictions based only on the Cygnus complex population.
\begin{table}[t]
\caption{Characteristics of the Cygnus OB associations and stellar clusters collectively referred to as the Cygnus complex \citep[from][]{Knodlseder:2002}.}
\begin{center}
\begin{tabular}{|c|c|c|c|}
\hline
\celltspace \cellbspace Name & Observed population & Distance (pc) & Age (Myr)\\
\hline
\celltspace \cellbspace Cyg OB1 & 23 $\in$ [15,40\,M$_{\odot}$] & 1905 & 4 \\
\hline
\celltspace \cellbspace Cyg OB2 & 120 $\in$ [20,120\,M$_{\odot}$] & 1584 & 2.5 \\
\hline
\celltspace \cellbspace Cyg OB3 & 14 $\in$ [25,60\,M$_{\odot}$] & 2187 & 3.5 \\
\hline
\celltspace \cellbspace Cyg OB7 & 10 $\in$ [7,25\,M$_{\odot}$] & 832 & 3.5 \\
\hline
\celltspace \cellbspace Cyg OB8 & 6 $\in$ [20,40\,M$_{\odot}$] & 2399 & 7.5 \\
\hline
\celltspace \cellbspace Cyg OB9 & 8 $\in$ [20,40\,M$_{\odot}$] & 1259 & 3.5 \\
\hline
\celltspace \cellbspace Ber 86 & 11 $\in$ [7,25\,M$_{\odot}$] & 1660 & 4 \\
\hline
\celltspace \cellbspace Ber 87 & 24 $\in$ [7,25\,M$_{\odot}$] & 1905 & 4.5 \\
\hline
\celltspace \cellbspace NGC 6871 & 13 $\in$ [7,25\,M$_{\odot}$] & 2399 & 5.5 \\
\hline
\celltspace \cellbspace NGC 6913 & 13 $\in$ [7,40\,M$_{\odot}$] & 1820 & 3.5 \\
\hline
\celltspace \cellbspace NGC 6910 & 7 $\in$ [7,25\,M$_{\odot}$] & 1820 & 4.5 \\
\hline
\celltspace \cellbspace NGC 6883 & 2 $\in$ [12,15\,M$_{\odot}$] & 1820 & 15 \\
\hline
\celltspace \cellbspace IC 4996 & 6 $\in$ [6,25\,M$_{\odot}$] & 1660 & 5.5 \\
\hline
\end{tabular}
\end{center}
\label{tab_ OBasso}
\end{table}

% The code
\subsection{The code}
\label{popsim_code}

\indent Our population synthesis code is very similar to many other such tools \citep[like][for example]{Leitherer:1992,Cervino:2000,Voss:2009}. For the case we are interested in, the parameters of a simulation are the characteristics of the stellar population to be modelled: slope and bounds of the IMF (initial mass function, assumed here to be a homogeneous power-law function), stellar content (given as the number of stars observed in a certain mass interval), and mean distance to the cluster (to compute the decay fluxes). We assume that the star formation occurs as an instantaneous burst, which means that all stars form at the same time (discussed later).\\
\indent Once the population to be synthesised has been defined, we create a cluster realisation by randomly sampling the IMF until the observational criterion (the number of observed stars) is met. Then, all stars are evolved in time steps typically of 10$^4$\,yrs. For each run, we compute the time-dependent release of $^{26}$Al and $^{60}$Fe, the decay fluxes at 1809 and 1173/1332\,keV and a few other outputs such as the mechanical luminosity. To evaluate the effects of finite sampling, a given stellar population is modelled several times (typically 100-200) until the mean stellar mass and number of stars obtained over all trials are within 1\% of the theoretical values for the specified IMF. For each quantity, this allows computing the mean value and the statistical variance at each time step in the life of the cluster.\\
\indent Although a synthetic star can take any initial mass between 11 and 120\,M$_{\odot}$ (the bounds of our grid of yields), we have the nucleosynthesis yields for only a finite number of initial stellar masses, typically 11, 12, 13, 14, 15, 16, 17, 20, 40, 60, 80, 120\,M$_{\odot}$ (the \citetalias{Palacios:2005} grids of the higher mass models differ for the various metallicities). We therefore need an interpolation scheme to compute the yields of the synthetic stars from the ones at our disposal.\\
\indent The characteristics of a synthetic star with a random initial mass are determined from the closest lower and higher mass models (for instance, the characteristics of a 53.6\,M$_{\odot}$ star are determined from the 40 and 60\,M$_{\odot}$ models). A first cubic-spline interpolation is performed to determine the durations of the main evolutionary sequences of the star: central H-burning, central He-burning and final stages (when available). Then, for each of these three sequences, the $^{26}$Al and $^{60}$Fe quantities ejected over a given time fraction of the sequence are obtained by linear interpolation of the quantities ejected by the closest lower and higher mass models over the same time fraction. When the synthetic star explodes, its $^{26}$Al and $^{60}$Fe yields are determined by a linear interpolation based on the CO core mass and released in the current time step.\\
\indent The work of \citetalias{Limongi:2006} clearly shows that the yields are not smooth functions of the initial stellar mass\footnote{On the contrary, the durations of the main burning stages are a quite smooth function of the initial mass, close to a power-law, and that is why in our work they are determined by a cubic-spline interpolation instead of a simple linear interpolation.}. For instance, the lack of a C convective shell in their 14 and 15\,M$_{\odot}$ models translates into a dramatic increase in the $^{26}$Al production compared to the 13 or 16\,M$_{\odot}$ models. As a consequence, it should be remembered that the finite grid of yields and the associated interpolation scheme are potential sources of errors.

% Predictions versus observations
\section{Predictions versus observations}
\label{predvsobs}

\indent Based on the code and inputs presented in Sect. \ref{popsim}, we have simulated the history of the ejection of $^{26}$Al and $^{60}$Fe in the Cygnus complex and computed the corresponding decay fluxes. We first focused on the purely photometric aspect, trying to reproduce the observed fluxes by simply adding the contributions from all the stellar groups listed in Table \ref{tab_ OBasso}. In the second part, we also attempted to reproduce the spatial extent of this emission.

% The decay emission fluxes
\subsection{The decay emission fluxes}
\label{predvsobs_fluxes}

\indent All the stellar groups listed in Table \ref{tab_ OBasso} have been modelled independently, assuming a coeval star formation for each of them. Then, the lightcurves at 1809 and 1173/1332\,keV were added, with the appropriate time shifts to take the differences in age of the different stellar populations into account. The resulting lightcurves for the whole Cygnus complex are given with t=0 corresponding to the present day and thus allowed exploring the past and future evolution of the decay fluxes.\\
\indent As a first step, we focused on the results obtained for an initial solar metallicity, firstly because most of the past population synthesis studies were based on nucleosynthesis yields from solar metallicity stellar models so a comparison could be made, but also because the Cygnus complex lies on the solar circle and is thus expected to have a nearly solar metallicity.

% Solar metallicity
\subsubsection{The case for solar metallicity Z=0.02}

\indent The results obtained for an initial solar metallicity are presented in Fig. \ref{fig_predvsobs} (top), together with the observational constraints we derived from the INTEGRAL/SPI observations. The predicted decay fluxes at 1809 and 1173/1332\,keV are there together with their typical variances due to the finite IMF sampling. From the open clusters and OB associations ages determined by \citet{Knodlseder:2002}, our current position along the time axis is indicated by the zero point. Yet, these cluster ages were determined by fitting isochrones to Hertzprung-Russel diagrams, and the isochrones used by these authors were built from the non-rotating stellar models available at that time. Rotation is now known to increase the lifetime of the models by 15-25\%, mainly through augmenting the main-sequence duration \citep{Meynet:2003}. Therefore, the cluster ages have probably been underestimated by the same amount, and our current position on the time axis of Fig. \ref{fig_predvsobs} is likely shifted to the right of the zero-point, as indicated by the grey shaded area.\\
\indent Inside this time range, there is clear agreement between the predicted 1809\,keV flux from $^{26}$Al and the observed value, while the theoretical level of the $^{60}$Fe decay emission is consistent with our upper limit. Our population synthesis indicates that we are currently in a period of steep increase of the 1809\,keV flux (over Myr timescales) that follows from a substantial release of $^{26}$Al through the stellar winds of the most massive stars, supplemented by the first supernova explosions that occurred in the oldest stellar groups, typically those above 3.5\,Myr (the shortest lifetime in our grid of models, corresponding to an initial 120\,M$_{\odot}$ star). At about 1\,Myr on the time axis of Fig. \ref{fig_predvsobs} (which may be our current position), most of the wind $^{26}$Al has been injected in the ISM; the supernovae then take over and now release both $^{26}$Al and $^{60}$Fe. From there on, the 1809\,keV flux is expected to rise again, less rapidly though, and to reach within $\sim$ 1\,Myr a maximum mean value of $\sim$ 4.2 $\times$ 10$^{-5}$\,ph\,cm$^{-2}$\,s$^{-1}$, which remains in our measured flux interval. Regarding the 1173/1332\,keV emission, the most massive stars of the oldest stellar groups exploded and consequently started injecting $^{60}$Fe in the ISM. The number of events up to now, however, is too low to have built up a detectable amount of $^{60}$Fe. Moreover, our prediction shows that even the maximum 1173/1332\,keV flux, to be reached in $\sim$ 2-3\,Myr from now, remains below our present upper limit.\\
\indent Our simulation predicts that a certain number of SNe, typically 10-20 over the last Myr, exploded in the Cygnus complex, which contradicts the apparent absence of radio SNRs and the detection of only a single pulsar in its various OB associations and open clusters. The absence of SNRs could be explained by the fact that SNRs expanding inside the hot tenuous interiors of superbubbles exhibit very weak characteristic radio and optical signatures \citep{Chu:1997}. With their high stellar content, the OB associations of the Cygnus complex are expected to blow large superbubbles inside of which the SNR are likely to be invisible. Moreover, the typical radio lifetime of a SNR is short compared to the lifetime of $^{26}$Al \citep[10$^4$ versus 10$^6$\,yrs,][]{Frail:1994} and all SNRs may well be completely diluted in the ISM by now. In that case, we would expect to see some manifestations of their compact remnants. Up to now, only a single pulsar has been proposed associated with the Cygnus complex, more precisely with Cyg OB2 \citep{Camilo:2009}, despite the dedicated deep surveys that have been undertaken at radio frequencies \citep{Janssen:2009}. The difficulty of finding pulsars in the Cygnus complex could come from an excessive scintillation or dispersion of the radio signals in the strongly ionised environment of such a concentration of massive stars. Or it may be that the narrow cones of their radio emission do not intersect our line of sight. In the latter case, these pulsars, if they exist, may become ``visible" to the Fermi/LAT instrument through their gamma-ray emission, which is unabsorbed and less-focused compared to the radio emission. The discovery by Fermi/LAT of the gamma-ray pulsar PSR J2032+4127 that is likely located in Cyg OB2 is promising in this respect \citep{Abdo:2009d}.\\
\indent The above result constitutes a step forward compared to the previous studies of the nucleosynthesis activity of the Cygnus region. Indeed, the latest evaluations of the situation conclude that, from the established stellar content (including the revised 100-120 O stars of Cyg OB2) and the stellar yields of \citet{Meynet:1997} and \citet{Woosley:1995}, the observed 1809\,keV flux from Cygnus is a factor of 2-3 above the predictions \citep[see for instance][]{Knodlseder:2002}. \citet{Pluschke:2000} point out that the possible enhanced yields of massive close binary systems could alleviate the discrepancy. While it is clear that such systems may occur in the Cygnus complex and could indeed have a strong impact on the nucleosynthesis yields, our theoretical knowledge of this scenario is still too uncertain to allow us to take it quantitatively into consideration. Following the significant upward revision of the Cyg OB2 stellar content \citep{Knodlseder:2000}, \citet{Pluschke:2002} suggested that the stellar content of the other OB associations of the Cygnus region may be underestimated due to a strong interstellar extinction; with correction factors based on the CO column densities in the direction of Cygnus, the authors show that the resulting increase in the stellar content could solve the problem, and yet, this revision of the stellar population has not been observationally substantiated since then. Moreover, \citet{Schneider:2006} show that some fraction of the CO mass observed toward Cygnus is actually located behind the OB associations, which means that the applied correction was incorrect in some cases.\\
\indent Our work indicates that the past dilemma resulted from an overestimate of the 1809\,keV flux of the Cygnus complex, combined with an underestimate of the nucleosynthesis yields. First of all, in \citetalias{Martin:2009a}, a careful analysis of the INTEGRAL/SPI 1809\,keV observations showed that the decay flux attributable to the Cygnus complex is only $\sim$ 65\% of the total flux coming from the Cygnus region. This result was obtained by separating the emission due to the Cygnus complex from the foreground and background mean Galactic contribution. Then, stellar rotation, as discussed in Sect. \ref{popsim_models}, turned out to be a key factor in nucleosynthesis since it enhances the production of $^{26}$Al but also affects the rate of its release in the ISM. In addition, the computation in a coherent way of the SNIb/c contribution (by \citetalias{Limongi:2006}) showed that the SN yields of the most massive stars cannot be neglected or simply extrapolated from pure He stars calculations.

% Cygnus metallicity
\subsubsection{The case for metallicity Z=0.01}

\begin{figure}[!t]
\begin{center}
\includegraphics[width=\columnwidth]{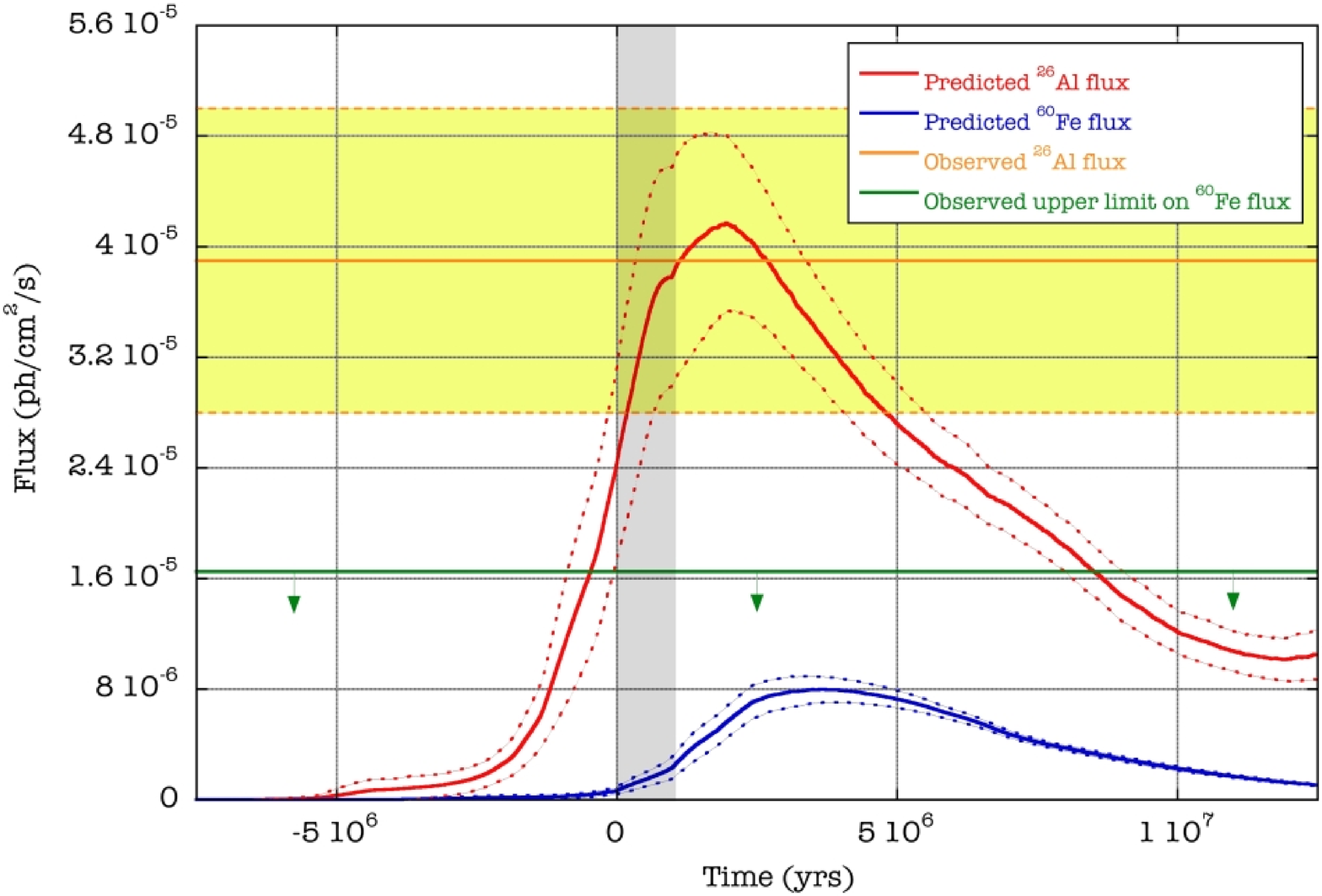}
\includegraphics[width=\columnwidth]{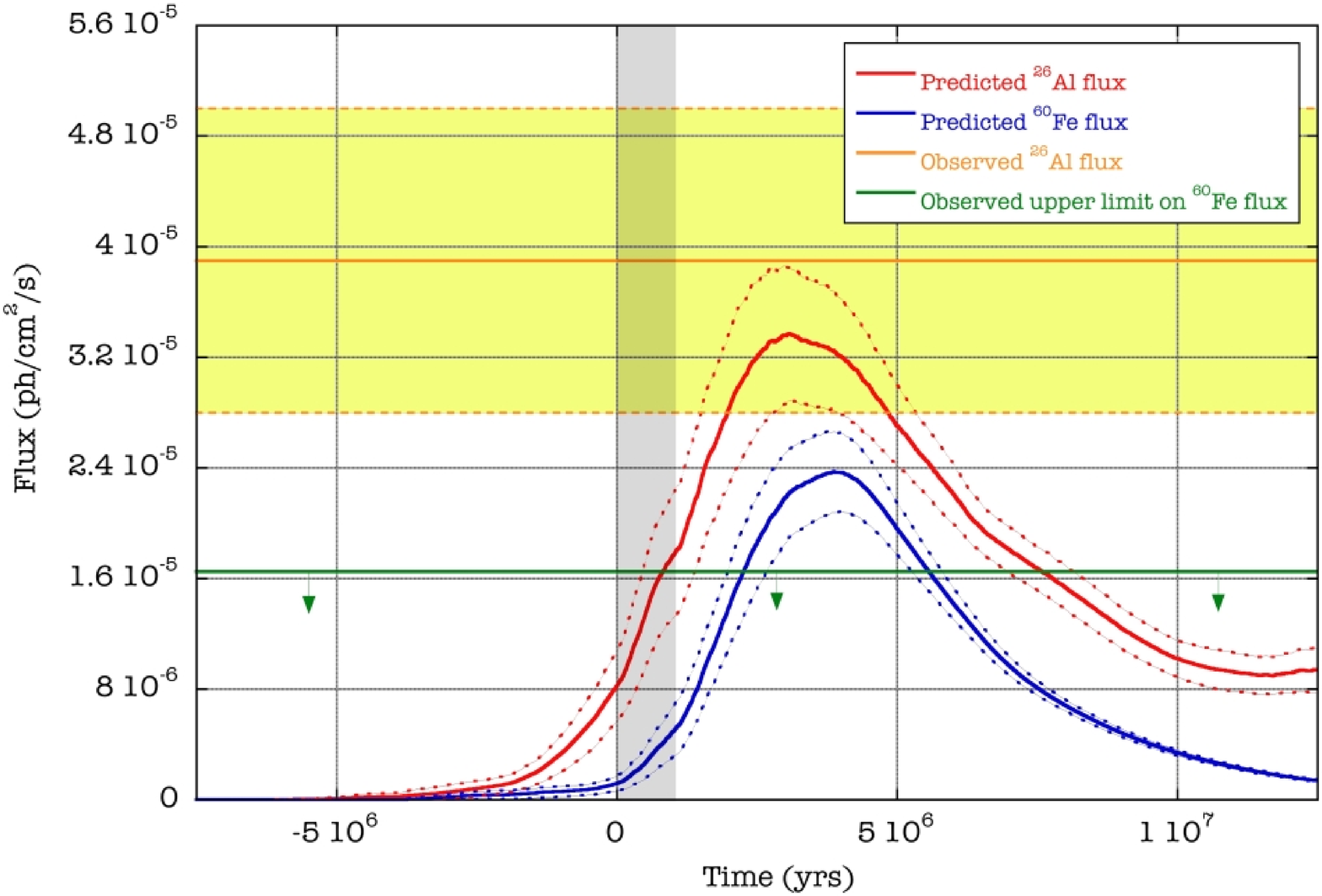}
\caption{Predicted versus observed decay fluxes, for Z=0.02 (solar) initial metallicity (top) and Z=0.01 initial metallicity (bottom). The red curve corresponds to the 1809\,keV flux and the blue curve to the 1173/1332\,keV flux. The dotted lines give the typical variance due to the finite IMF sampling. The orange band indicates the observed 1809\,keV flux and the green line the upper limit on the 1173/1332\,keV emission from the Cygnus complex. The vertical shaded area indicates our probable current position along the time axis (see text).}
\label{fig_predvsobs}
\end{center}
\end{figure}

\indent Most of the past studies were based on the nucleosynthesis yields of solar metallicity stellar models, so we presented the above result for comparison; yet, there are indications that the stellar population of the Cygnus region may be subsolar with a metallicity of about Z=0.01 \citep{Daflon:2001}. Although we do not have a fully homogeneous grid of stellar models of various metallicities covering hydrostatic and explosive phases at our disposal, we exploited as much as possible the works of \citetalias{Limongi:2006} and \citetalias{Palacios:2005} to estimate the $^{26}$Al and $^{60}$Fe yields of a non-solar population of massive stars. We recall here that our grid of yields was built from purely hydrostatic stellar tracks of various initial metallicities combined with full calculations (hydrostatic and explosive) only at solar metallicity, the connection being made (when necessary) through the CO core mass (see \ref{popsim_grid}).\\
\indent The decay lightcurves for Z=0.01 (obtained by linear interpolation between Z=0.008 and Z=0.02) are presented in Fig. \ref{fig_predvsobs} (bottom). Compared to the Z=0.02 case, the peak 1809\,keV flux is somewhat reduced and occurs at a later time, while the 1173/1332\,keV flux is increased by almost a factor of 3. Both differences come from the effect of lower metallicity on stellar evolution. Indeed, a lower initial metallicity implies reduced mass loss for stars initially more massive than 20\,M$_{\odot}$, since the winds of the OB and WR phases are mostly driven by radiative pressure on metals through their numerous absorption lines. Reduced mass loss leads to larger CO core masses, hence higher SN yields, as we assumed that the latter mostly depends on the CO core mass. This accounts for the considerable increase in the $^{60}$Fe production in the Z=0.01 case. For $^{26}$Al, however, the increase in the SN yields is compensated by a decrease in the wind yields; indeed, a lower initial metallicity means a reduced mass loss but also a reduced initial $^{25}$Mg reservoir for proton capture (hence synthesis of $^{26}$Al) during central H-burning. On the whole, the 1809\,keV lightcurve is not dramatically altered by the reduced metallicity, but the stronger SN contribution explains the delay in the 1809\,keV peak flux compared to the Z=0.02 case.\\
\indent Within the likely present-day time window, the mean predicted 1809\,keV flux is at best a factor of two below the observed value. Consistency is found later on, at $\sim$ 2-3\,Myr from now, but then the 1173/1332\,keV flux has risen close to or even above our upper limit. It therefore seems that the overall agreement obtained at Z=0.02 does no longer holds if the subsolar metallicity of Cygnus is accounted for. We may question this subsolar metallicity. It is indeed somewhat surprising that stellar associations lying at about the same galactocentric distance as the Sun and being much younger could be subsolar, although \citet{Daflon:2001} mention that their abundances are consistent with the expected values, as predicted by various Galactic abundance gradients. While the C, N O, Si, Al, and Fe were found to be underabundant relative to solar values, Mg and S were found to be solar and Mg has straightforward relevance for $^{26}$Al nucleosynthesis. The authors state that B stars in the solar neighbourhood generally have subsolar abundances, so one may worry about a general systematic bias in determining chemical abundances in B stars. We also note that a solar abundance was derived by \citet{Najarro:2001} for the P-Cygni star, which lies at $\sim$ 2\,kpc, close to the Cygnus complex.\\
\indent From these arguments, we therefore consider that the metallicity of the Cygnus complex is still uncertain, especially since the associations investigated by \citet{Daflon:2001} are Cyg OB3 and 7, which are not those contributing the most to the Cygnus complex. In addition to this, we discuss below several other potential sources of error or uncertainty that can modify and/or alleviate the current situation for the gamma-ray line emission of the Cygnus complex.

% Sources of error/uncertainty
\subsubsection{Sources of error/uncertainty}
\label{predvsobs_fluxes_error}

\begin{figure}[t]
\begin{center}
\includegraphics[width=\columnwidth]{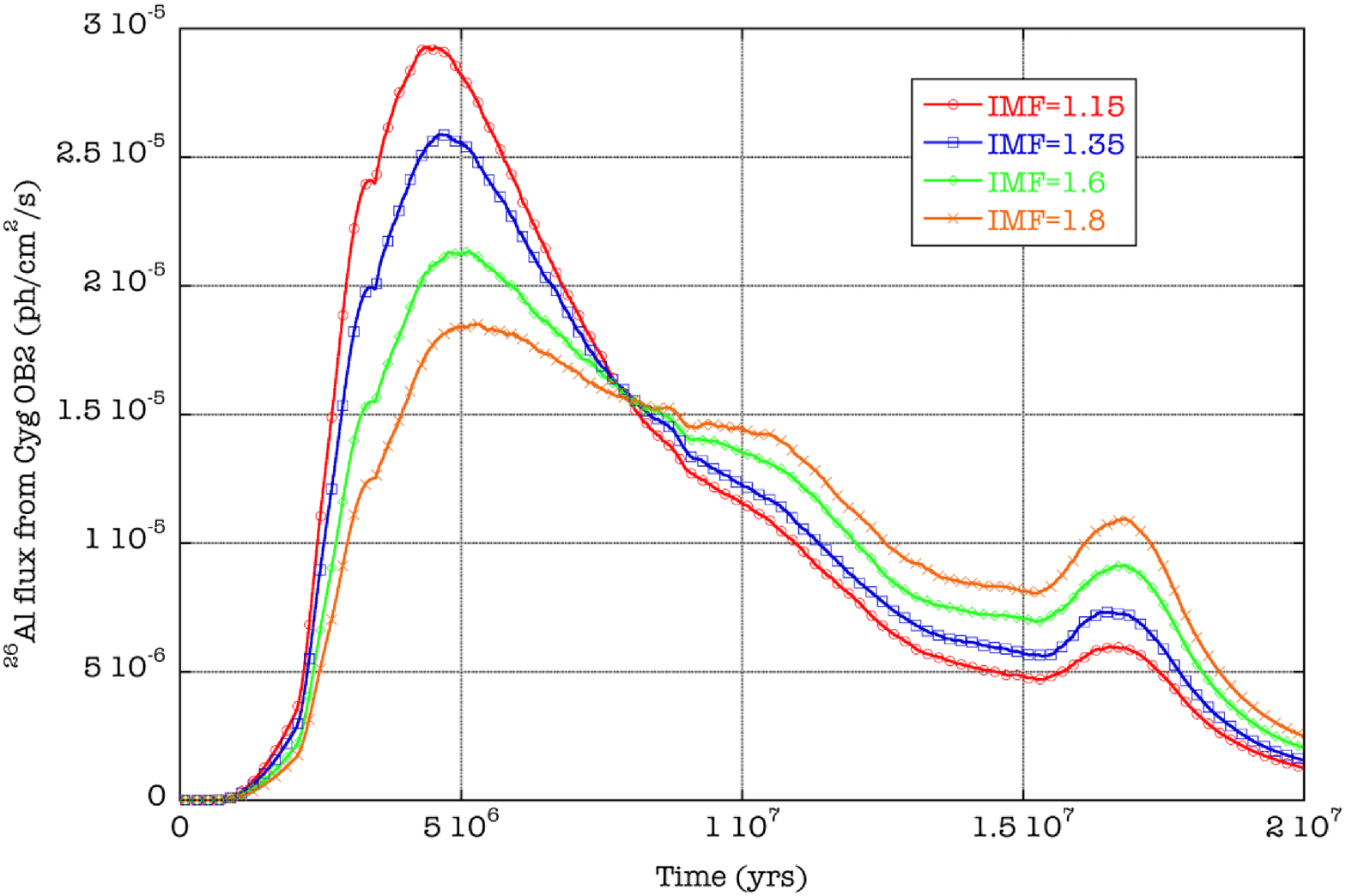}
\includegraphics[width=\columnwidth]{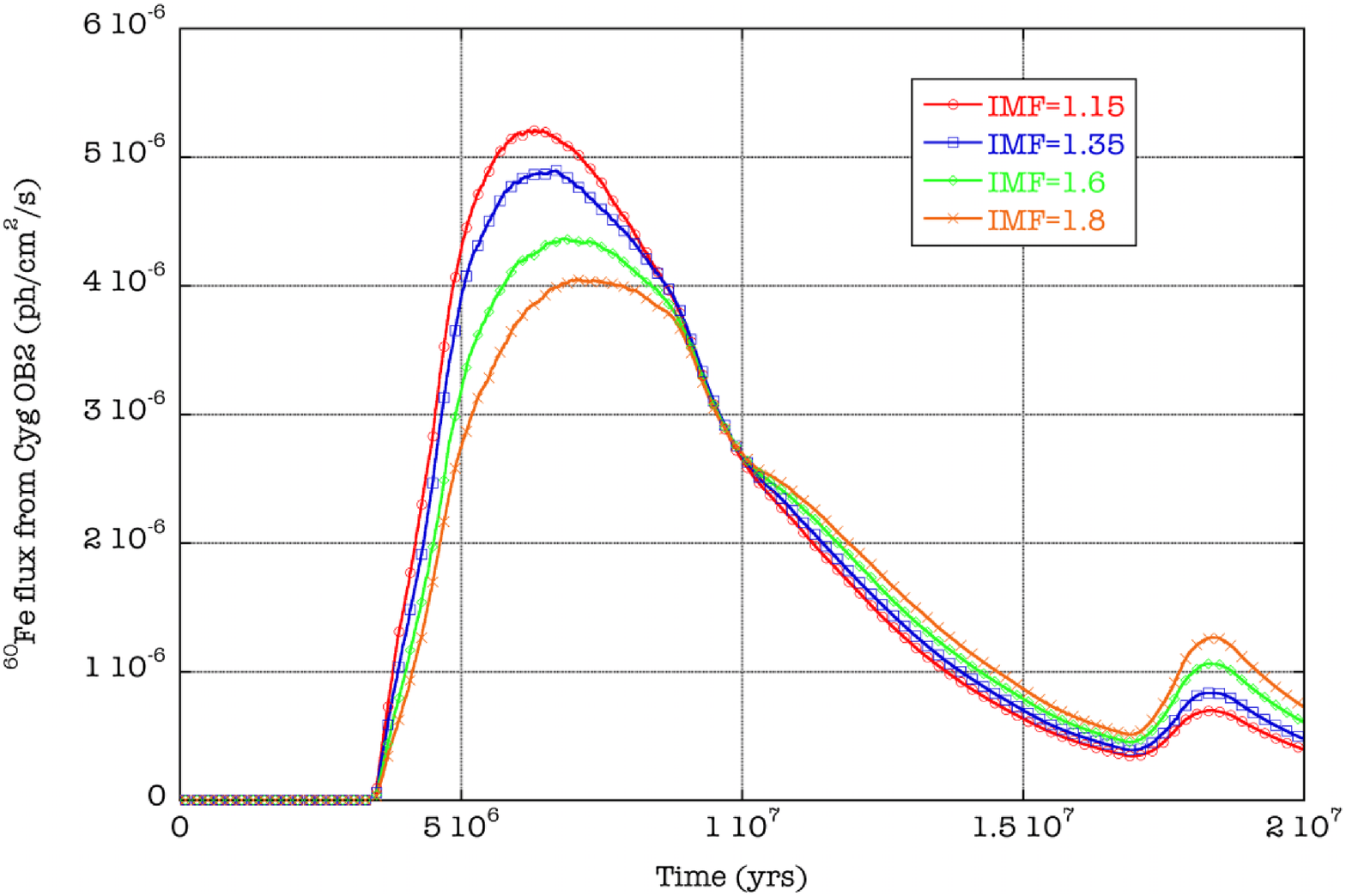}
\caption{Decay fluxes of $^{26}$Al and $^{60}$Fe from Cyg OB2 as a function of the IMF slope for an initial solar metallicity. In contrast to Fig. \ref{fig_predvsobs}, the t=0 point does not mark the present day but instead the date of the coeval star formation.}
\label{fig_fluxvsIMF}
\end{center}
\end{figure}

\indent Up to now, we have only considered the uncertainty on the predicted fluxes due to finite IMF sampling, but the uncertainty on the exact distances and ages of the stellar associations of the Cygnus complex should also be taken into account. From \citet{Knodlseder:2002}, we estimated that around the present date, the uncertainties on the ages and distances of the stellar clusters translate into typical uncertainties of $\sim$ 1.0 and 0.25 $\times$ 10$^{-5}$\,ph\,cm$^{-2}$\,s$^{-1}$ on the decay line fluxes of $^{26}$Al and $^{60}$Fe, respectively. In connection with the age of the stellar clusters, it should be noted that our assumption of a coeval star formation favours maximum peak fluxes. The most massive stars (between 30 and 120\,M$_{\odot}$) have very similar lifetimes, so in a coeval population the nucleosynthesis contributions of their last stages (WR and SNe) are released at almost the same time and therefore lead to a steep rise in the decay fluxes. In contrast, if star formation occurs over a few Myrs, the nucleosynthesis contributions are spread over a longer time interval and the resulting peak fluxes are lower.\\
\indent We emphasise once again that our grid of nucleosynthesis yields is not homogeneous and self-consistent and that the SN yields of the higher-mass stars at all metallicities were extrapolated from non-rotating solar models (whereas the hydrostatic phases of these higher mass stars were computed by including stellar rotation). As a first potential issue, the direct effect of metallicity on the SN yields of non-solar models is not taken into account. Indeed, our grid of yields takes into account the effect of metallicity on stellar evolution processes like the mass loss and the late stage core sizes. But the direct influence of metallicity, that is the differing chemical composition of the stellar matter from which the nucleosynthesis proceeds, is not considered. For instance, we have shown before that an initial subsolar metallicity leads to bigger CO core, hence presumably to higher $^{60}$Fe SN yields; however, this should be mitigated by the fact that a lower metallicity also means reduced initial quantities of Fe and CNO for the synthesis of $^{60}$Fe. Since $^{60}$Fe is a pure secondary element, it is likely that our grid overestimates the $^{60}$Fe SN yields of the subsolar metallicity models. Regarding $^{26}$Al, the direct impact of the metallicity is less straightforward. Indeed, most of the $^{26}$Al released by supernovae is of explosive origin and the corresponding nuclear processes seem to depend only mildly on the initial metallicity (although the many involved species do not allow drawing definitive conclusions). In this case, our original assumption that the SN yields predominantly depend on the CO core mass appears to be more reliable for $^{26}$Al than for $^{60}$Fe.\\
\indent A second potential issue stems from SN yields coming from non-rotating models. Rotation has proven to be efficient at enhancing the $^{26}$Al production during central H-burning, mostly thanks to an increased convection and mixing. If rotation were to modify the yields of the C/Ne shell-burning episodes in a similar manner, the corresponding $^{26}$Al and $^{60}$Fe production would be higher than those computed by \citetalias{Limongi:2006}. In addition, increasing the shell convection efficiency would also alter the final mass-radius relation on which our estimate of the explosive nucleosynthesis depends. We should also mention that, in the grid of yields we are using, the $\nu$-process\footnote{In the $\nu$-process, protons liberated by the $\nu$-spallation of e.g. $^{20}$Ne, $^{16}$O, $^{23}$Na, and $^{24}$Mg capture on $^{25}$Mg to form $^{26}$Al.} has not been taken into account. This production channel was estimated to enhance the explosive $^{26}$Al yields by 30-50\% \citep{Woosley:1995,Timmes:1995}. These values may have to be revised with more sophisticated and realistic neutrino spectra and transport.\\
\indent More generally, the nucleosynthesis yields should be regarded as sensitive to most ingredients of stellar models, such as convection and mixing, opacities or nuclear reaction rates. In \ref{popsim_grid}, we compared the \citetalias{Limongi:2006} and \citetalias{Woosley:2007} sets of yields and showed that the choice of mass loss prescription or the implementation of convective physics does significantly affect the nucleosynthesis of $^{26}$Al and $^{60}$Fe. We did not observe drastic variations by many orders of magnitude in the outputs of the two grids of stellar models, but the typical differences are high enough to possibly modify our conclusions. Therefore, although they were not included in our population synthesis, we qualitatively assessed the impact that the \citetalias{Woosley:2007} yields would have on the predicted decay fluxes from the Cygnus complex.\\
\indent For stars in the 20-120\,M$_{\odot}$ mass range (which dominate the current nucleosynthesis activity of the Cygnus complex), the $^{26}$Al wind and SN yields of the \citetalias{Woosley:2007} models are clearly lower than those of \citetalias{Palacios:2005} and \citetalias{Limongi:2006} (see Figs. \ref{fig_26Alwindyields} and \ref{fig_26AlSNyields}). We estimated that the use of these yields as such in our population synthesis would lower the 1809\,keV flux by at least a factor of 2 and thus lead to a mismatch between predictions and observations if the Cygnus complex has nearly solar metallicity. Nevertheless, the \citetalias{Woosley:2007} models do not include stellar rotation, which turned out to strongly enhance the $^{26}$Al yields. Then, keeping the hopefully more realistic yields of \citetalias{Palacios:2005} for the $^{26}$Al wind contribution, the lower $^{26}$Al SN yields of the \citetalias{Woosley:2007} models are not expected to seriously alter the rising part of the early 1809\,keV lightcurve that we observe today (see Fig. \ref{fig_predvsobs}), because the latter is dominated by the wind contribution. This is clearly apparent in Fig. \ref{fig_fluxvsIMF} for the simulation of Cyg OB2 only: when the first stars of the coeval cluster explode, at $\sim$3.5\,Myr, the flux has already risen to 70-80\% of its maximum value and is already consistent with the observed value. Consequently, even if the \citetalias{Woosley:2007} models may lead to less $^{26}$Al being ejected by SNe, the wind yields of the presently available rotating models alone can account for most of the observed $^{26}$Al decay flux. Regarding $^{60}$Fe, the \citetalias{Woosley:2007} yields are not very different from the yields we have been using (\citetalias{Palacios:2005} models extrapolated through CO core mass from \citetalias{Limongi:2006}) over the 20-120\,M$_{\odot}$ mass range (see Fig. \ref{fig_60FeSNyields}). We thus expect the 1173/1332\,keV lightcurve to still be consistent with our upper limit event if the \citetalias{Woosley:2007} yields were used.\\
\indent We therefore believe that the \citetalias{Woosley:2007} set of yields would not modify our results for the solar metallicity case if we still use the enhanced $^{26}$Al wind yields of the rotating models. A definitive conclusion would require fully implementing \citetalias{Woosley:2007} in our population synthesis code (in order, for instance, to take the specific SN yields to CO core mass relation of this grid properly into account). More generally, as alternative sets of yields become available, especially from rotating models, it will be interesting to see if our conclusions still hold. Even in the absence of major improvements of the stellar models, the evolution of some of their inputs alone, such as nuclear data, can alter the nucleosynthesis results. Some critical reaction rates, such as $\alpha$(2$\alpha$,$\gamma$)$^{12}$C or $^{12}$C($\alpha$,$\gamma$)$^{16}$O, are known to have strong effects on the final yields \citep[see for instance][]{Tur:2009}. As an illustration of the uncertainties on nuclear data, the half-life of $^{60}$Fe was recently revised to 2.6\,Myr, instead of the 1.5\,Myr value we have been using throughout this work \citep{Rugel:2009}. In terms of predicted 1173/1332\,keV lightcurves for the Cygnus complex, however, this longer lifetime simply implies an even lower emission at early times, still consistent with our upper limit \citep[for a more quantitative analysis of the impact of the new $^{60}$Fe lifetime, see][]{Voss:2009}.\\
\indent We also emphasise that all the results presented so far are based on a ``classical" IMF slope of 1.35 \citep{Salpeter:1955}. Since the universality of this parameter is still a matter of debate, we show in Fig. \ref{fig_fluxvsIMF} how the decay lightcurves for Cyg OB2 vary with the slope of the IMF. The impact is fairly straightforward: a flatter IMF means more very massive stars (for a given total number of massive stars) and the 1809\,keV peak at 3-4\,Myr attributable to the winds and SNe of WR stars is consequently stronger, while a steeper IMF means fewer very massive stars, which reduces the 1809\,peak flux at 3-4\,Myr and enhances the flux at 15-16\,Myr, as a result of more SN explosions of 13-15\,M$_{\odot}$ stars.\\
\indent In connection with the potential effect of the IMF slope, we want to draw attention to the possible impact of binarity. It is commonly accepted that for stellar clusters more distant than a few kpc, most of the observed point-sources assumed to be members of the clusters may actually be unresolved binaries or multiple systems. The census of their stellar content is consequently biased, particularly for purely photometric approaches, but even when relying on spectroscopy. \citet{Kobulnicky:2007} find that the binary fraction of Cyg OB2 is very high, and may well be close to 1; moreover, it seems that massive stars preferentially have massive companions. We therefore wonder about the possibility that a substantial fraction of the 100-120 O stars of Cyg OB2, many of which have been identified photometrically from the 2MASS data, may in fact be unresolved massive binaries or multiple systems. If this were true, the number of massive stars in our population synthesis would have to be increased, which could lead to higher predicted fluxes. To properly account for this binarity effect, however, the various observations devoted to Cyg OB2 (and the other OB associations) have to be thoroughly compared to clearly identify which fraction of the objects assumed today as single massive stars may actually hide multiples, but this is beyond the scope of the present paper.

% The spatial diffusion of 26Al
\subsection{The spatial diffusion of $^{26}$Al}
\label{predvsobs_distrib}

\begin{figure}[t]
\begin{center}
\includegraphics[width=\columnwidth]{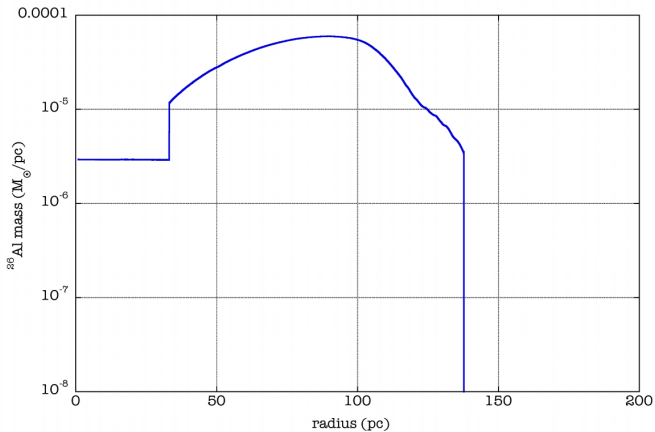}
\caption{Distribution of $^{26}$Al inside the superbubble blown by Cyg OB2 after 3\,Myr, as predicted from a simple numerical simulation.}
\label{fig_26AlmassSBdistrib}
\end{center}
\end{figure}

In the previous sections, we made use of the photometric aspects of the INTEGRAL/SPI observations of Cygnus alone. Here we want to extend our interpretation of the SPI results to the spatial and spectrometric information provided by this instrument.\\
\indent The numerous massive stars of the Cygnus complex are expected to have blown one or several superbubbles (SBs) through the combined action of their winds and SN explosions \citep{Castor:1975,Weaver:1977,Tomisaka:1986,Mac-Low:1988,Chu:2004,Chu:2008}. Soft X-ray observations by HEAO-3 and ROSAT have revealed a large structure in the Cygnus direction that may well be the outermost part of a gigantic SB, the centre of which is concealed by intervening interstellar material \citep{Cash:1980}. Fresh $^{26}$Al is expected to be injected inside the hot and tenuous interior of this SB, and its diffusion on large scales will then be driven by the growth of that structure.\\
\indent Under certain simplifications, a self-similar analytical solution can be found for the growth of the SB \citep{Castor:1975}:
\begin{align}
\label{eq_sbana}
& R_{s} = 66 \times L_{38}^{1/5} n_{1}^{-1/5} t_{6}^{3/5} \quad \textrm{pc} \\
& V_{s} = 40 \times L_{38}^{1/5} n_{1}^{-1/5} t_{6}^{-2/5} \quad \textrm{km} \, \textrm{s} ^{-1}
\end{align}
where $R_{s}$ and $V_{s}$ are the outer radius and velocity of the SB, respectively, given as functions of the mechanical luminosity of the central cluster $L_{38}$ (in 10$^{38}$\,erg\,s$^{-1}$), the uniform density of the surrounding medium $n_{1}$ (in cm$^{-3}$) and the age of the central cluster $t_{6}$ (in Myr). These equations will allow first estimates of the expected size of the 1809\,keV diffuse emission from Cygnus, but also show the relatively weak dependence of the bubble's size on some uncertain parameters, such as the density of the surrounding ISM medium.\\
\indent In the following, we focus on the Cyg OB2 cluster only because we expect most of the mechanical power and nucleosynthesis activity of the Cygnus complex to come from it. From our population synthesis tool, we have found that the mechanical luminosity of Cyg OB2 is about 4 $\times$ 10$^{38}$\,erg\,s$^{-1}$ over the first 3\,Myr of the cluster's life\footnote{More details about the calculation of the mechanical power can be found in \citet{Voss:2009}, who also developed a population synthesis code dedicated to gamma-ray line flux predictions. Their code is based on similar assumptions and the same set of stellar data, and the results of both tools have been checked to be consistent.}. Using this value, together with a typical ISM density of 1\,cm$^{-3}$, we find that the radius of the SB blown by Cyg OB2 is about 170\,pc after 3\,Myr. Assuming that the $^{26}$Al homogeneously fills the interior of the SB, the corresponding 1809\,keV decay signal would appear to us as a diffuse emission with a typical size of 12$^{\circ}$ (given the $\sim$1.6\,kpc distance of Cyg OB2). This estimate tallies with our INTEGRAL/SPI measurement of a 9-10$^{\circ}$ angular size (especially since the maximum size of the emission is weakly constrained by SPI, see \ref{obs_26Al}). It is also consistent with previous measurements by CGRO/COMPTEL \citep{Pluschke:2001}.\\
\indent We advanced the accuracy of our estimate of the expected extent of the 1809\,keV diffuse emission from Cygnus by using a numerical simulation. Our observational constraints are actually not tight enough to justify such an effort, but the present study was an opportunity to address the question of the diffusion of $^{26}$Al around a stellar cluster in some details. We used the VH-1 hydrodynamical code\footnote{Made available to the community by John Blondin and collaborators, see http://wonka.physics.ncsu.edu/pub/VH-1} to compute the evolution of a simple spherical bubble blown by the injection of kinetic energy at the centre. The code was modified to include an equation for the advection and decay of $^{26}$Al in addition to the three classical Euler equations. A variable input was also implemented to allow injection of $^{26}$Al according to the time profile given by our population synthesis code. Last, radiative losses, which rapidly become significant in the swept-up ISM shell, were implemented in the form of a cooling function from \citet{Sutherland:1993}.\\
\indent By again using a constant mechanical luminosity of 4 $\times$ 10$^{38}$\,erg\,s$^{-1}$ with a typical wind velocity of 2000\,km\,s$^{-1}$, and an ISM density of 1\,cm$^{-3}$ for a medium with 90\% H and 10\% He (in number), the resulting $^{26}$Al mass distribution after about 3\,Myr is plotted in Fig. \ref{fig_26AlmassSBdistrib}. The inner flat distribution corresponds to the rapid advection with the freely-expanding winds, followed by a sharp discontinuity that marks the inner boundary of the shocked-wind region that actually constitutes most of the interior of the SB. A sudden drop in the $^{26}$Al content then occurs at the contact discontinuity that separates the shocked stellar material from the shocked swept-up ISM. From Fig. \ref{fig_26AlmassSBdistrib}, it appears that the $^{26}$Al fills the entire SB but somehow accumulates at roughly half its radius. The $^{26}$Al mass distribution is relatively homogeneous inside the hot shocked interior of the SB, with variations along the radius that remain within a factor of 10.\\
\indent According to our numerical simulation, the angular size of the 1809\,keV emission from the $^{26}$Al released by the Cyg OB2 cluster is $\sim$ 10$^{\circ}$, which is again consistent with the measurements by INTEGRAL/SPI and CGRO/COMPTEL. Our simulation also indicates that most of the $^{26}$Al is affected by expansion velocities in the 50-300\,km\,s$^{-1}$ range, which agrees with the velocities inferred from the 1809\,keV line broadening. We want to caution, however, that several physical processes were neglected in our numerical model. First of all, thermal conduction between the hot tenuous interior of the SB and the cold dense swept-up ISM shell was not taken into account \citep{Castor:1975,Weaver:1977}. This energy transfer causes material from the shell to evaporate and enrich the mass content of the SB (the resulting mass flow actually dominates the mass carried by the stellar winds). Nevertheless, \citet{Tomisaka:1986}, show that this process does not significantly alter the overall size of the structure, but rather acts as an internal mass-energy redistribution. Second, our one-dimensional simulation cannot account for the turbulence that very likely occurs in the SB as a result of the multiple wind flows inside the stellar cluster, inhomogeneities in the surrounding medium or hydrodynamical instabilities. Such turbulence will mix the stellar material inside the bubble and probably ablate mass from the swept-up ISM shell. These two effects will obviously blur the contact discontinuity and alter the mass distribution compared to what we presented in Fig. \ref{fig_26AlmassSBdistrib}. Turbulent motions will also be an additional source for the broadening of the decay line.

% Conclusion
\section{Conclusion}
\label{conclusion}

We have compared the INTEGRAL/SPI observations of the gamma-ray line emission of the young OB associations and open clusters of the Cygnus region with theoretical expectations based on the latest models of stellar nucleosynthesis. This nearby complex of massive stars exhibits an unambiguous diffuse emission at 1809\,keV caused by the decay of $^{26}$Al but no detectable signal of the decay of $^{60}$Fe at 1173/1332\,keV. The latest studies of the nucleosynthesis activity of the Cygnus region show that the 1809\,keV emission deduced from CGRO/COMPTEL observations was a factor of 2-3 above the predictions from the theoretical yields available then and the observed stellar content of the Cygnus region.\\
\indent Compared to previous works, our grid of yields includes some of the effects of stellar rotation for the higher-mass stars and a coherent estimate of the contribution from SNIb/c. From this, we show that the observed decay fluxes from the Cygnus complex are consistent with the values predicted by population synthesis at solar metallicity. Our work indicates that the past dilemma resulted from overestimating the 1809\,keV flux of the Cygnus complex, combined with underestimating the nucleosynthesis yields. We showed in a previous work, through a careful analysis of the INTEGRAL/SPI 1809\,keV observations, that the decay flux attributable to the Cygnus complex is only $\sim$ 65\% of the total flux coming from the Cygnus region. This result was obtained by separating the emission due to the Cygnus complex from the foreground and background mean Galactic contribution. Then, stellar rotation turned out to be a key factor in nucleosynthesis since it enhances the production of $^{26}$Al but also affects the rate of its release in the ISM. Last, the coherent computation of the SNIb/c contribution (by \citetalias{Limongi:2006}) has led to an upward revision of the SN yields of the most massive stars.\\
\indent Nonetheless, when extrapolated to the subsolar metallicity that is indicated by some observations of the Cygnus complex, our predictions fail to account for the INTEGRAL/SPI measurements. Since no complete and coherent grid of nucleosynthesis yields for non-solar metallicity stellar models is available today, the discrepancy may well come from our extrapolation from the currently available data. One could also question, however, the metallicity estimate and the fact that a stellar population lying at about the same galactocentric distance as the Sun and being much younger could be subsolar. This discrepancy will not be solved until complete nucleosynthesis calculations at non-solar metallicities are performed and/or until the subsolar metallicity of the Cygnus complex is confirmed by a more recent estimate using improved stellar atmosphere models. These works are all the more expected because the solar metallicity has recently been revised from Z=0.02 to Z=0.014, which will obviously affect nucleosynthesis \citep{Asplund:2005}. In the meantime, our current understanding of massive stars and their nucleosynthesis will be tested further by applying the analysis presented here to other nearby OB associations with detectable gamma-ray line emission, like Sco-Cen. The potential effect of binarity on the predicted nucleosynthesis activity should also be evaluated.\\
\indent Finally, we performed a numerical simulation of the diffusion of $^{26}$Al inside a superbubble. Applied to the massive Cyg OB2 cluster, our model predicts that after 3\,Myr, the $^{26}$Al fills the entire superbubble and extends up to a radius of $\sim$ 140\,pc, in agreement with the observed extent of the 1809\,keV emission from Cygnus. Our simulation also indicates that most of the $^{26}$Al is affected by expansion velocities in the 50-300\,km\,s$^{-1}$ range, which agrees with the velocities inferred from the 1809\,keV line broadening.

\begin{acknowledgement}
The SPI project has been completed under the responsibility and leadership of CNES. We are grateful to the ASI, CEA, CNES, DLR, ESA, INTA, NASA, and OSTC for their support. Pierrick Martin wishes to thank Marco Limongi \& Alessandro Chieffi who kindly prepared and provided the nucleosynthesis data of their stellar models. 
\end{acknowledgement}

\bibliographystyle{aa}
\bibliography{/Users/pierrickmartin/Documents/MyPapers/biblio/Cygnus&CygOB2,/Users/pierrickmartin/Documents/MyPapers/biblio/Nucleosynthesis,/Users/pierrickmartin/Documents/MyPapers/biblio/SNobservations,/Users/pierrickmartin/Documents/MyPapers/biblio/SPI,/Users/pierrickmartin/Documents/MyPapers/biblio/26Al&60Fe,/Users/pierrickmartin/Documents/MyPapers/biblio/SNRobservations,/Users/pierrickmartin/Documents/MyPapers/biblio/GalaxyObservations,/Users/pierrickmartin/Documents/MyPapers/biblio/44Ti,/Users/pierrickmartin/Documents/MyPapers/biblio/ScoCen,/Users/pierrickmartin/Documents/MyPapers/biblio/StellarModels,/Users/pierrickmartin/Documents/MyPapers/biblio/Superbubbles,/Users/pierrickmartin/Documents/MyPapers/biblio/Pulsars,/Users/pierrickmartin/Documents/MyPapers/biblio/IMF,/Users/pierrickmartin/Documents/MyPapers/biblio/Fermi}

\begin{thebibliography}{63}
\expandafter\ifx\csname natexlab\endcsname\relax\def\natexlab#1{#1}\fi

\bibitem[{{Abdo} {et~al.}(2009){Abdo}, {Ackermann}, {Ajello}, {Anderson},
  {Atwood}, {Axelsson}, {Baldini}, {Ballet}, {Barbiellini}, {Baring},
  {Bastieri}, {Baughman}, {Bechtol}, {Bellazzini}, {Berenji}, {Bignami},
  {Blandford}, {Bloom}, {Bonamente}, {Borgland}, {Bregeon}, {Brez}, {Brigida},
  {Bruel}, {Burnett}, {Caliandro}, {Cameron}, {Caraveo}, {Casandjian},
  {Cecchi}, {{\c C}elik}, {Chekhtman}, {Cheung}, {Chiang}, {Ciprini}, {Claus},
  {Cohen-Tanugi}, {Conrad}, {Cutini}, {Dermer}, {de Angelis}, {de Luca}, {de
  Palma}, {Digel}, {Dormody}, {do Couto e Silva}, {Drell}, {Dubois}, {Dumora},
  {Farnier}, {Favuzzi}, {Fegan}, {Fukazawa}, {Funk}, {Fusco}, {Gargano},
  {Gasparrini}, {Gehrels}, {Germani}, {Giebels}, {Giglietto}, {Giommi},
  {Giordano}, {Glanzman}, {Godfrey}, {Grenier}, {Grondin}, {Grove},
  {Guillemot}, {Guiriec}, {Gwon}, {Hanabata}, {Harding}, {Hayashida}, {Hays},
  {Hughes}, {J{\'o}hannesson}, {Johnson}, {Johnson}, {Johnson}, {Kamae},
  {Katagiri}, {Kataoka}, {Kawai}, {Kerr}, {Kn{\"o}dlseder}, {Kocian}, {Kuss},
  {Lande}, {Latronico}, {Lemoine-Goumard}, {Longo}, {Loparco}, {Lott},
  {Lovellette}, {Lubrano}, {Madejski}, {Makeev}, {Marelli}, {Mazziotta},
  {McConville}, {McEnery}, {Meurer}, {Michelson}, {Mitthumsiri}, {Mizuno},
  {Monte}, {Monzani}, {Morselli}, {Moskalenko}, {Murgia}, {Nolan}, {Norris},
  {Nuss}, {Ohsugi}, {Omodei}, {Orlando}, {Ormes}, {Paneque}, {Parent},
  {Pelassa}, {Pepe}, {Pesce-Rollins}, {Pierbattista}, {Piron}, {Porter},
  {Primack}, {Rain{\`o}}, {Rando}, {Ray}, {Razzano}, {Rea}, {Reimer}, {Reimer},
  {Reposeur}, {Ritz}, {Rochester}, {Rodriguez}, {Romani}, {Ryde},
  {Sadrozinski}, {Sanchez}, {Sander}, {Parkinson}, {Scargle}, {Sgr{\`o}},
  {Siskind}, {Smith}, {Smith}, {Spandre}, {Spinelli}, {Starck}, {Strickman},
  {Suson}, {Tajima}, {Takahashi}, {Takahashi}, {Tanaka}, {Thayer}, {Thompson},
  {Tibaldo}, {Tibolla}, {Torres}, {Tosti}, {Tramacere}, {Uchiyama}, {Usher},
  {Van Etten}, {Vasileiou}, {Vilchez}, {Vitale}, {Waite}, {Wang}, {Watters},
  {Winer}, {Wolff}, {Wood}, {Ylinen}, \& {Ziegler}}]{Abdo:2009d}
{Abdo}, A.~A., {Ackermann}, M., {Ajello}, M., {et~al.} 2009, Science, 325, 840

\bibitem[{{Asplund} {et~al.}(2005){Asplund}, {Grevesse}, \&
  {Sauval}}]{Asplund:2005}
{Asplund}, M., {Grevesse}, N., \& {Sauval}, A.~J. 2005, in Astronomical Society
  of the Pacific Conference Series, Vol. 336, Cosmic Abundances as Records of
  Stellar Evolution and Nucleosynthesis, ed. T.~G. {Barnes}, III \& F.~N.
  {Bash}, 25--+

\bibitem[{{Camilo} {et~al.}(2009){Camilo}, {Ray}, {Ransom}, {Burgay},
  {Johnson}, {Kerr}, {Gotthelf}, {Halpern}, {Reynolds}, {Romani}, {Demorest},
  {Johnston}, {van Straten}, {Saz Parkinson}, {Ziegler}, {Dormody}, {Thompson},
  {Smith}, {Harding}, {Abdo}, {Crawford}, {Freire}, {Keith}, {Kramer},
  {Roberts}, {Weltevrede}, \& {Wood}}]{Camilo:2009}
{Camilo}, F., {Ray}, P.~S., {Ransom}, S.~M., {et~al.} 2009, \apj, 705, 1

\bibitem[{{Cash} {et~al.}(1980){Cash}, {Charles}, {Bowyer}, {Walter},
  {Garmire}, \& {Riegler}}]{Cash:1980}
{Cash}, W., {Charles}, P., {Bowyer}, S., {et~al.} 1980, \apjl, 238, L71

\bibitem[{{Castor} {et~al.}(1975){Castor}, {McCray}, \& {Weaver}}]{Castor:1975}
{Castor}, J., {McCray}, R., \& {Weaver}, R. 1975, \apjl, 200, L107

\bibitem[{{Cervi{\~n}o} {et~al.}(2000){Cervi{\~n}o}, {Kn{\"o}dlseder},
  {Schaerer}, {von Ballmoos}, \& {Meynet}}]{Cervino:2000}
{Cervi{\~n}o}, M., {Kn{\"o}dlseder}, J., {Schaerer}, D., {von Ballmoos}, P., \&
  {Meynet}, G. 2000, \aap, 363, 970

\bibitem[{{Chu}(1997)}]{Chu:1997}
{Chu}, Y.-H. 1997, \aj, 113, 1815

\bibitem[{{Chu}(2008)}]{Chu:2008}
{Chu}, Y.-H. 2008, in IAU Symposium, Vol. 250, IAU Symposium, 341--354

\bibitem[{{Chu} {et~al.}(2004){Chu}, {Guerrero}, \& {Gruendl}}]{Chu:2004}
{Chu}, Y.-H., {Guerrero}, M.~A., \& {Gruendl}, R.~A. 2004, in Astrophysics and
  Space Science Library, Vol. 315, How Does the Galaxy Work?, ed. E.~J.
  {Alfaro}, E.~{P{\'e}rez}, \& J.~{Franco}, 165--+

\bibitem[{{Comer{\'o}n} \& {Pasquali}(2007)}]{Comeron:2007}
{Comer{\'o}n}, F. \& {Pasquali}, A. 2007, \aap, 467, L23

\bibitem[{{Comer{\'o}n} {et~al.}(2008){Comer{\'o}n}, {Pasquali}, {Figueras}, \&
  {Torra}}]{Comeron:2008}
{Comer{\'o}n}, F., {Pasquali}, A., {Figueras}, F., \& {Torra}, J. 2008, \aap,
  486, 453

\bibitem[{{Comer{\'o}n} {et~al.}(2002){Comer{\'o}n}, {Pasquali}, {Rodighiero},
  {Stanishev}, {De Filippis}, {L{\'o}pez Mart{\'{\i}}}, {G{\'a}lvez Ortiz},
  {Stankov}, \& {Gredel}}]{Comeron:2002}
{Comer{\'o}n}, F., {Pasquali}, A., {Rodighiero}, G., {et~al.} 2002, \aap, 389,
  874

\bibitem[{{Daflon} {et~al.}(2001){Daflon}, {Cunha}, {Becker}, \&
  {Smith}}]{Daflon:2001}
{Daflon}, S., {Cunha}, K., {Becker}, S.~R., \& {Smith}, V.~V. 2001, \apj, 552,
  309

\bibitem[{{Dame} {et~al.}(2001){Dame}, {Hartmann}, \& {Thaddeus}}]{Dame:2001}
{Dame}, T.~M., {Hartmann}, D., \& {Thaddeus}, P. 2001, \apj, 547, 792

\bibitem[{{del Rio} {et~al.}(1996){del Rio}, {von Ballmoos}, {Bennett},
  {Bloemen}, {Diehl}, {Hermsen}, {Knoedlseder}, {Oberlack}, {Ryan},
  {Schoenfelder}, \& {Winkler}}]{Del-Rio:1996}
{del Rio}, E., {von Ballmoos}, P., {Bennett}, K., {et~al.} 1996, \aap, 315, 237

\bibitem[{{Diehl} {et~al.}(2006){Diehl}, {Halloin}, {Kretschmer}, {Lichti},
  {Sch{\"o}nfelder}, {Strong}, {von Kienlin}, {Wang}, {Jean}, {Kn{\"o}dlseder},
  {Roques}, {Weidenspointner}, {Schanne}, {Hartmann}, {Winkler}, \&
  {Wunderer}}]{Diehl:2006}
{Diehl}, R., {Halloin}, H., {Kretschmer}, K., {et~al.} 2006, \nat, 439, 45

\bibitem[{{Diehl} {et~al.}(1994){Diehl}, {Kn{\"o}dlseder}, {Lichti},
  {Sch{\"o}nfelder}, {Steinle}, {Strong}, {Dupraz}, {Bloemen}, {Hermsen},
  {Swanenburg}, {Morris}, {Ryan}, {Stacy}, {Bennett}, \&
  {Winkler}}]{Diehl:1994}
{Diehl}, R., {Kn{\"o}dlseder}, J., {Lichti}, G., {et~al.} 1994, in American
  Institute of Physics Conference Series, Vol. 304, American Institute of
  Physics Conference Series, ed. C.~E. {Fichtel}, N.~{Gehrels}, \& J.~P.
  {Norris}, 147--155

\bibitem[{{Frail} {et~al.}(1994){Frail}, {Goss}, \& {Whiteoak}}]{Frail:1994}
{Frail}, D.~A., {Goss}, W.~M., \& {Whiteoak}, J.~B.~Z. 1994, \apj, 437, 781

\bibitem[{{Harris} {et~al.}(2005){Harris}, {Kn{\"o}dlseder}, {Jean}, {Cisana},
  {Diehl}, {Lichti}, {Roques}, {Schanne}, \& {Weidenspointner}}]{Harris:2005}
{Harris}, M.~J., {Kn{\"o}dlseder}, J., {Jean}, P., {et~al.} 2005, \aap, 433,
  L49

\bibitem[{{Hunter} {et~al.}(2008){Hunter}, {Brott}, {Lennon}, {Langer},
  {Dufton}, {Trundle}, {Smartt}, {de Koter}, {Evans}, \& {Ryans}}]{Hunter:2008}
{Hunter}, I., {Brott}, I., {Lennon}, D.~J., {et~al.} 2008, \apjl, 676, L29

\bibitem[{{Iwamoto} {et~al.}(1999){Iwamoto}, {Brachwitz}, {Nomoto},
  {Kishimoto}, {Umeda}, {Hix}, \& {Thielemann}}]{Iwamoto:1999}
{Iwamoto}, K., {Brachwitz}, F., {Nomoto}, K., {et~al.} 1999, \apjs, 125, 439

\bibitem[{{Janssen} {et~al.}(2009){Janssen}, {Stappers}, {Braun}, {van
  Straten}, {Edwards}, {Rubio-Herrera}, {van Leeuwen}, \&
  {Weltevrede}}]{Janssen:2009}
{Janssen}, G.~H., {Stappers}, B.~W., {Braun}, R., {et~al.} 2009, \aap, 498, 223

\bibitem[{{Kaplan} {et~al.}(2004){Kaplan}, {Frail}, {Gaensler}, {Gotthelf},
  {Kulkarni}, {Slane}, \& {Nechita}}]{Kaplan:2004}
{Kaplan}, D.~L., {Frail}, D.~A., {Gaensler}, B.~M., {et~al.} 2004, \apjs, 153,
  269

\bibitem[{{Karakas} \& {Lattanzio}(2007)}]{Karakas:2007}
{Karakas}, A. \& {Lattanzio}, J.~C. 2007, Publications of the Astronomical
  Society of Australia, 24, 103

\bibitem[{{Kn{\"o}dlseder}(2000)}]{Knodlseder:2000}
{Kn{\"o}dlseder}, J. 2000, \aap, 360, 539

\bibitem[{{Kn{\"o}dlseder} {et~al.}(1999){Kn{\"o}dlseder}, {Bennett},
  {Bloemen}, {Diehl}, {Hermsen}, {Oberlack}, {Ryan}, {Sch{\"o}nfelder}, \& {von
  Ballmoos}}]{Knodlseder:1999}
{Kn{\"o}dlseder}, J., {Bennett}, K., {Bloemen}, H., {et~al.} 1999, \aap, 344,
  68

\bibitem[{{Kn{\"o}dlseder} {et~al.}(2002){Kn{\"o}dlseder}, {Cervi{\~n}o}, {Le
  Duigou}, {Meynet}, {Schaerer}, \& {von Ballmoos}}]{Knodlseder:2002}
{Kn{\"o}dlseder}, J., {Cervi{\~n}o}, M., {Le Duigou}, J.-M., {et~al.} 2002,
  \aap, 390, 945

\bibitem[{{Kobulnicky} \& {Fryer}(2007)}]{Kobulnicky:2007}
{Kobulnicky}, H.~A. \& {Fryer}, C.~L. 2007, \apj, 670, 747

\bibitem[{{Langer}(1989)}]{Langer:1989}
{Langer}, N. 1989, \aap, 220, 135

\bibitem[{{Le Duigou} \& {Kn{\"o}dlseder}(2002)}]{Le-Duigou:2002}
{Le Duigou}, J.-M. \& {Kn{\"o}dlseder}, J. 2002, \aap, 392, 869

\bibitem[{{Leitherer} {et~al.}(1992){Leitherer}, {Robert}, \&
  {Drissen}}]{Leitherer:1992}
{Leitherer}, C., {Robert}, C., \& {Drissen}, L. 1992, \apj, 401, 596

\bibitem[{{Limongi} \& {Chieffi}(2006)}]{Limongi:2006}
{Limongi}, M. \& {Chieffi}, A. 2006, \apj, 647, 483

\bibitem[{{Lodders}(2003)}]{Lodders:2003}
{Lodders}, K. 2003, \apj, 591, 1220

\bibitem[{{Mac Low} \& {McCray}(1988)}]{Mac-Low:1988}
{Mac Low}, M.-M. \& {McCray}, R. 1988, \apj, 324, 776

\bibitem[{{Maeder} {et~al.}(2009){Maeder}, {Meynet}, {Ekstr{\"o}m}, \&
  {Georgy}}]{Maeder:2009}
{Maeder}, A., {Meynet}, G., {Ekstr{\"o}m}, S., \& {Georgy}, C. 2009,
  Communications in Asteroseismology, 158, 72

\bibitem[{{Manchester} {et~al.}(2005){Manchester}, {Hobbs}, {Teoh}, \&
  {Hobbs}}]{Manchester:2005}
{Manchester}, R.~N., {Hobbs}, G.~B., {Teoh}, A., \& {Hobbs}, M. 2005, \aj, 129,
  1993

\bibitem[{{Martin} {et~al.}(2009){Martin}, {Kn{\"o}dlseder}, {Diehl}, \&
  {Meynet}}]{Martin:2009a}
{Martin}, P., {Kn{\"o}dlseder}, J., {Diehl}, R., \& {Meynet}, G. 2009, \aap,
  506, 703

\bibitem[{{Meynet} {et~al.}(1997){Meynet}, {Arnould}, {Prantzos}, \&
  {Paulus}}]{Meynet:1997}
{Meynet}, G., {Arnould}, M., {Prantzos}, N., \& {Paulus}, G. 1997, \aap, 320,
  460

\bibitem[{{Meynet} \& {Maeder}(2003)}]{Meynet:2003}
{Meynet}, G. \& {Maeder}, A. 2003, \aap, 404, 975

\bibitem[{{Meynet} \& {Maeder}(2005)}]{Meynet:2005}
{Meynet}, G. \& {Maeder}, A. 2005, \aap, 429, 581

\bibitem[{{Najarro}(2001)}]{Najarro:2001}
{Najarro}, F. 2001, in Astronomical Society of the Pacific Conference Series,
  Vol. 233, P Cygni 2000: 400 Years of Progress, ed. M.~{de Groot} \&
  C.~{Sterken}, 133--+

\bibitem[{{Nugis} \& {Lamers}(2000)}]{Nugis:2000}
{Nugis}, T. \& {Lamers}, H.~J.~G.~L.~M. 2000, \aap, 360, 227

\bibitem[{{Palacios} {et~al.}(2005){Palacios}, {Meynet}, {Vuissoz},
  {Kn{\"o}dlseder}, {Schaerer}, {Cervi{\~n}o}, \& {Mowlavi}}]{Palacios:2005}
{Palacios}, A., {Meynet}, G., {Vuissoz}, C., {et~al.} 2005, \aap, 429, 613

\bibitem[{{Pl{\"u}schke} {et~al.}(2002){Pl{\"u}schke}, {Cervi{\~n}o}, {Diehl},
  {Kretschmer}, {Hartmann}, \& {Kn{\"o}dlseder}}]{Pluschke:2002}
{Pl{\"u}schke}, S., {Cervi{\~n}o}, M., {Diehl}, R., {et~al.} 2002, New
  Astronomy Review, 46, 535

\bibitem[{{Pl{\"u}schke} {et~al.}(2001){Pl{\"u}schke}, {Diehl},
  {Sch{\"o}nfelder}, {Bloemen}, {Hermsen}, {Bennett}, {Winkler}, {McConnell},
  {Ryan}, {Oberlack}, \& {Kn{\"o}dlseder}}]{Pluschke:2001}
{Pl{\"u}schke}, S., {Diehl}, R., {Sch{\"o}nfelder}, V., {et~al.} 2001, in ESA
  Special Publication, Vol. 459, Exploring the Gamma-Ray Universe, ed.
  A.~{Gimenez}, V.~{Reglero}, \& C.~{Winkler}, 55--58

\bibitem[{{Pl{\"u}schke} {et~al.}(2000){Pl{\"u}schke}, {Diehl},
  {Sch{\"o}nfelder}, {Weidenspointer}, {Bolemen}, {Hermsen}, {McConnell},
  {Ryan}, {Bennett}, {Oberlack}, \& {Kn{\"o}dlseder}}]{Pluschke:2000}
{Pl{\"u}schke}, S., {Diehl}, R., {Sch{\"o}nfelder}, V., {et~al.} 2000, in
  American Institute of Physics Conference Series, Vol. 510, American Institute
  of Physics Conference Series, ed. M.~L. {McConnell} \& J.~M. {Ryan}, 35--+

\bibitem[{{Prantzos} \& {Diehl}(1996)}]{Prantzos:1996}
{Prantzos}, N. \& {Diehl}, R. 1996, \physrep, 267, 1

\bibitem[{{Rugel} {et~al.}(2009){Rugel}, {Faestermann}, {Knie}, {Korschinek},
  {Poutivtsev}, {Schumann}, {Kivel}, {G{\"u}nther-Leopold}, {Weinreich}, \&
  {Wohlmuther}}]{Rugel:2009}
{Rugel}, G., {Faestermann}, T., {Knie}, K., {et~al.} 2009, Physical Review
  Letters, 103, 072502

\bibitem[{{Salpeter}(1955)}]{Salpeter:1955}
{Salpeter}, E.~E. 1955, \apj, 121, 161

\bibitem[{{Schneider} {et~al.}(2006){Schneider}, {Bontemps}, {Simon}, {Jakob},
  {Motte}, {Miller}, {Kramer}, \& {Stutzki}}]{Schneider:2006}
{Schneider}, N., {Bontemps}, S., {Simon}, R., {et~al.} 2006, \aap, 458, 855

\bibitem[{{Smith}(2004)}]{Smith:2004}
{Smith}, D.~M. 2004, in ESA Special Publication, Vol. 552, 5th INTEGRAL
  Workshop on the INTEGRAL Universe, ed. V.~{Schoenfelder}, G.~{Lichti}, \&
  C.~{Winkler}, 45--+

\bibitem[{{Sutherland} \& {Dopita}(1993)}]{Sutherland:1993}
{Sutherland}, R.~S. \& {Dopita}, M.~A. 1993, \apjs, 88, 253

\bibitem[{{Timmes} {et~al.}(1995){Timmes}, {Woosley}, {Hartmann}, {Hoffman},
  {Weaver}, \& {Matteucci}}]{Timmes:1995}
{Timmes}, F.~X., {Woosley}, S.~E., {Hartmann}, D.~H., {et~al.} 1995, \apj, 449,
  204

\bibitem[{{Tomisaka} \& {Ikeuchi}(1986)}]{Tomisaka:1986}
{Tomisaka}, K. \& {Ikeuchi}, S. 1986, \pasj, 38, 697

\bibitem[{{Tur} {et~al.}(2009){Tur}, {Heger}, \& {Austin}}]{Tur:2009}
{Tur}, C., {Heger}, A., \& {Austin}, S.~M. 2009, submitted to \apj

\bibitem[{{Voss} {et~al.}(2009){Voss}, {Diehl}, {Hartmann}, {Cervi{\~n}o},
  {Vink}, {Meynet}, {Limongi}, \& {Chieffi}}]{Voss:2009}
{Voss}, R., {Diehl}, R., {Hartmann}, D.~H., {et~al.} 2009, \aap, 504, 531

\bibitem[{{Wang} {et~al.}(2007){Wang}, {Harris}, {Diehl}, {Halloin}, {Cordier},
  {Strong}, {Kretschmer}, {Kn{\"o}dlseder}, {Jean}, {Lichti}, {Roques},
  {Schanne}, {von Kienlin}, {Weidenspointner}, \& {Wunderer}}]{Wang:2007}
{Wang}, W., {Harris}, M.~J., {Diehl}, R., {et~al.} 2007, \aap, 469, 1005

\bibitem[{{Weaver} {et~al.}(1977){Weaver}, {McCray}, {Castor}, {Shapiro}, \&
  {Moore}}]{Weaver:1977}
{Weaver}, R., {McCray}, R., {Castor}, J., {Shapiro}, P., \& {Moore}, R. 1977,
  \apj, 218, 377

\bibitem[{{Wellstein} \& {Langer}(1999)}]{Wellstein:1999}
{Wellstein}, S. \& {Langer}, N. 1999, \aap, 350, 148

\bibitem[{{Wendker} {et~al.}(1991){Wendker}, {Higgs}, \&
  {Landecker}}]{Wendker:1991}
{Wendker}, H.~J., {Higgs}, L.~A., \& {Landecker}, T.~L. 1991, \aap, 241, 551

\bibitem[{{Woosley} \& {Heger}(2007)}]{Woosley:2007}
{Woosley}, S.~E. \& {Heger}, A. 2007, \physrep, 442, 269

\bibitem[{{Woosley} {et~al.}(1995){Woosley}, {Langer}, \&
  {Weaver}}]{Woosley:1995a}
{Woosley}, S.~E., {Langer}, N., \& {Weaver}, T.~A. 1995, \apj, 448, 315

\bibitem[{{Woosley} \& {Weaver}(1995)}]{Woosley:1995}
{Woosley}, S.~E. \& {Weaver}, T.~A. 1995, \apjs, 101, 181

\end{thebibliography}

\end{document}